\documentclass[twocolumn,amssymb,pra,superscriptaddress,floatfix]{revtex4}

\usepackage{amsmath}
\usepackage{amsfonts}
\usepackage{graphicx}

\newcommand{\Ha}{\mathcal{H}}

\newcommand{\wid}{8.cm}
\newcommand{\wide}{16.2cm}

\begin{document}

\newcommand{\placeFigOne}{\begin{figure}[t!]
\centering
\flushleft{(a)}\\\includegraphics[width=\wid]{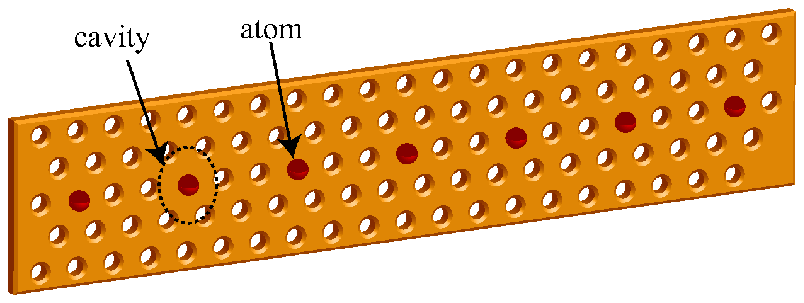}\\\vspace{0.1cm}
\flushleft{(b)}\\\includegraphics[width=\wid]{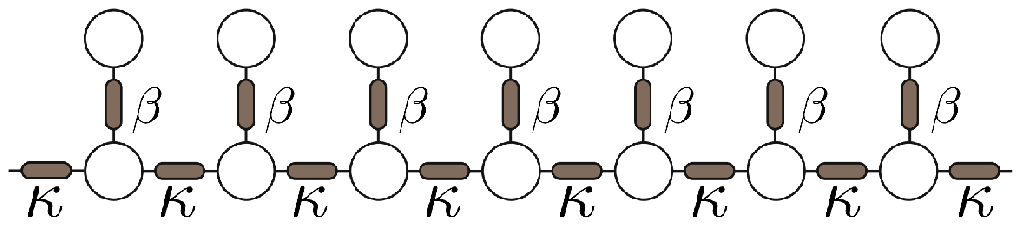}\\
\caption{(Color online) Visualizations of a 1D JCH system.  Fig.~(a)
  shows an example of a coupled cavity waveguide in a photonic
  crystal.  A lattice of holes in the membrane provide variations in
  refractive index, hence trapping photons of a given frequency.  The
  red spheres indicate two-level systems, these are placed at sites
  without holes, called defects, which form photonic
  cavities. Fig.~(b) is a visualization of a portion of a JCH chain as
  logical elements.  The bottom row of circles indicate the photonic
  cavities, linked together by the hopping rate $\kappa$. The top row
  of circles indicate the atoms, linked back to each photonic cavity
  by the coupling strength $\beta$ [defined in
    Eq.~(\ref{eq:mainham})].}
\label{fig:1DJCH}
\end{figure}
}

\newcommand{\placeFigTwo}{\begin{figure*}[th!]
\centering
\includegraphics[width=17cm]{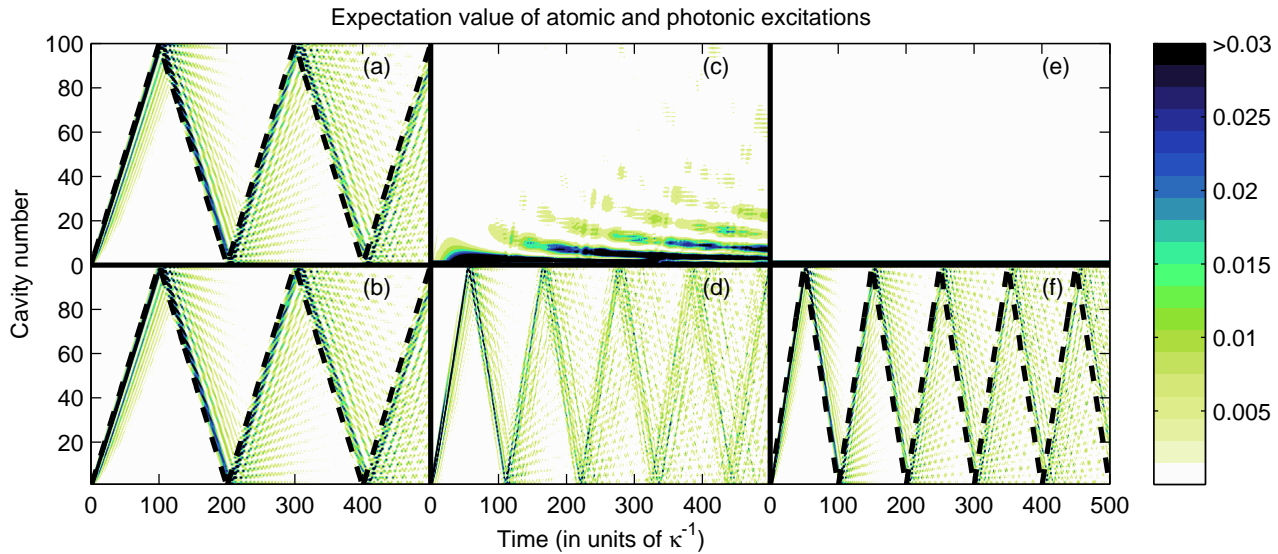}\\
\caption{ (Color online) Space-time diagrams for evolution of an
  excitation ($|1\rangle\otimes(|g,1\rangle+|e,0\rangle)/\sqrt{2}$)
  along a chain of 100 JC cavities for three different parameter
  regimes.  In each case, we plot the probability of occupation of a
  particular cavity (vertical axis) as a function of time (horizontal
  axis) for the system initialized in an even superposition of the
  atomic and photonic mode of the first cavity.  The upper three plots
  show the population of the atomic components whereas the lower three
  plots are the photonic components.  The ratio of cavity-cavity
  coupling to atom-photon coupling varies from left to right, with (a)
  and (b) $\kappa/\beta=10^{-3}$, (c) and (d) $\kappa/\beta=10$ and
  (e) and (f) $\kappa/\beta=10^3$.  In plots (a), (b) and (f), the
  dashed lines represent $\mathcal{Q}_{\wedge}$, as given in
  Eq.~(\ref{eq:triangle}).  }
\label{fig:bigUniformBottom}
\end{figure*}
}

\newcommand{\placeFigThree}{\begin{figure*}[th!]  
\centering
\includegraphics[width=\wide]{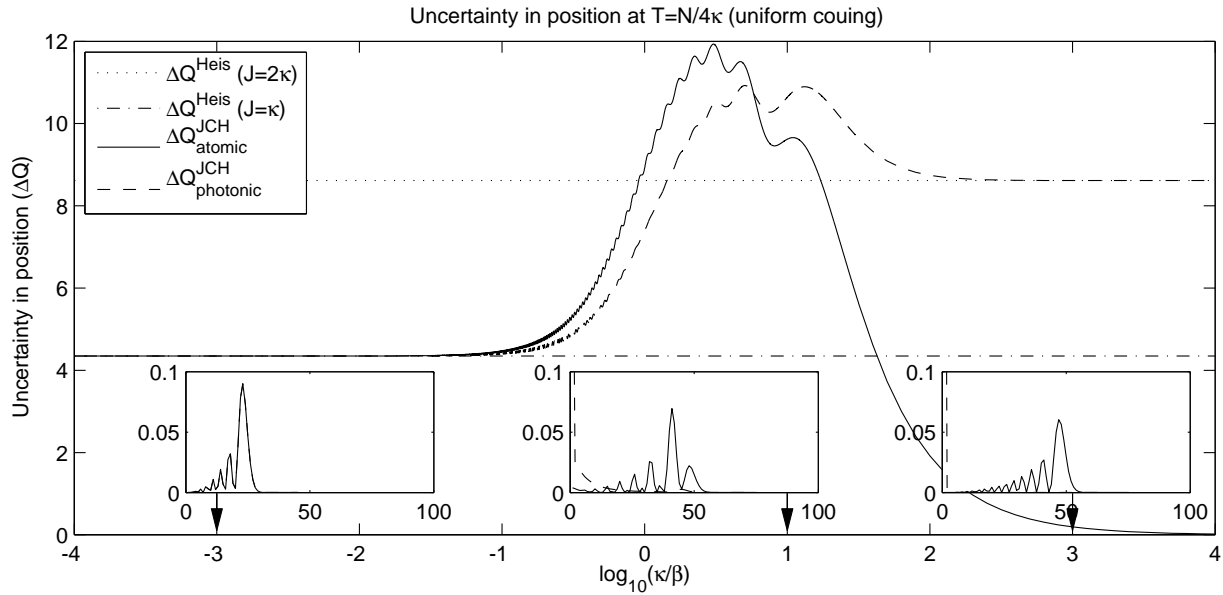}
\caption{ Dispersion of the wave packet, $\Delta Q$, at a fixed point
  in time, as a function of $\kappa/\beta$, with atom-cavity detuning
  $\Delta=0$ and number of cavities $N=100$.  The solid (dashed) line
  shows $\Delta Q_{\rm atomic}^{\rm JCH}$ ($\Delta Q_{\rm
    photonic}^{\rm JCH}$).  The dotted (dot-dashed) line shows $\Delta
  Q^{\rm Heis}$ with $J=2\kappa$ ($J=\kappa$). The dispersion is
  measured at a time $T=N/4\kappa$, at which point the packet is
  `freely' evolving along the chain.  This point is chosen such that
  the effects of the hard-wall boundaries can be ignored for both
  large and small $\kappa$. At the left of the plot,
  $\kappa/\beta\ll1$, and the JCH chain mimics two identical
  Heisenberg spin chains, an example is shown of this localized
  behavior in Fig.~(\ref{fig:bigUniformBottom})(a) and (b), where
  $\kappa/\beta=10^{-3}$.  The corresponding spatial profile of the
  pulse at $T=N/4\kappa$ is shown in the left inset (the solid line
  shows the photonic profile and the dashed line shows the atomic
  profile, in this case they are coincident).  At the right of the
  plot, $\kappa/\beta\gg 1$, and the photonic mode of the JCH chain
  mimics a Heisenberg spin chain, while the atomic mode does not
  propagate at all. An example of this type of localized behavior is
  shown in Fig.~(\ref{fig:bigUniformBottom})(e) and (f), where
  $\kappa/\beta=10^3$, the corresponding profile of the pulse at
  $T=N/4\kappa$ is shown in the right inset.  In the middle of the
  plot a travelling excitation shows a large amount of dispersion, an
  example of this delocalized behavior is shown in
  Fig.~(\ref{fig:bigUniformBottom})(c) and (d), where
  $\kappa/\beta=10$, and the corresponding profile at $T=N/4\kappa$ is
  shown in the middle inset.  The horizontal lines show the dispersion
  $\Delta Q^{\rm Heis}$ of a Heisenberg spin chain with 100 spins
  after the initial state $|1\rangle$ evolves until the front of the
  excitation is half way (dotted line) and quarter way (dot-dashed
  line) along the excitation chain.  The exact correspondence between
  the Heisenberg and JCH systems is seen in these asymptotic limits.}
\label{fig:bigUniformTop}
\end{figure*}
}

\newcommand{\placeFigFour}{\begin{figure}[t!]
\centering
\includegraphics[height=6.8cm]{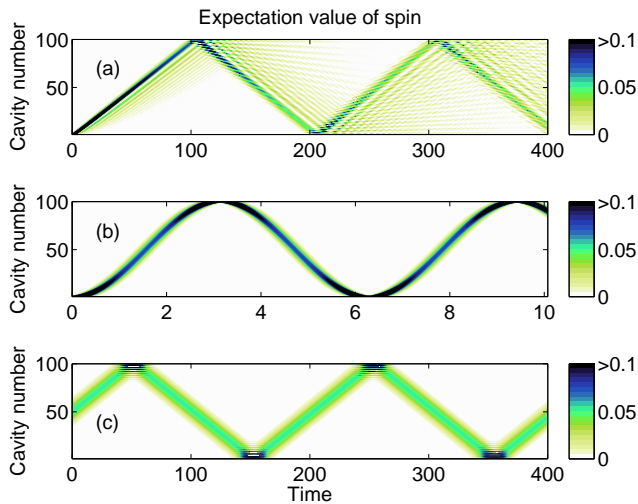}\\
\caption{(Color online) Evolution of a Heisenberg spin chain
  Hamiltonian [Eq.~(\ref{eq:Heis1up})], with $J=1$.  (a) Evolution of
  $|1\rangle$ under uniform coupling, note that at first the
  excitation travels fairly neatly, but both within a single pass, as
  well as during the reflection, the excitation pulse spreads out.
  Note the faint lines which are parallel but displaced from the
  wavefront.  These lines are due to the nature of the Heisenberg
  Hamiltonian.  (b) Evolution of $|1\rangle$ under parabolic coupling,
  note how the evolution is smooth and repetitive, the excitation
  travels sinusoidally from one end of the line of spins to the other
  end and back again, without spreading out. (c) Evolution of a
  Gaussian pulse as in Eq.~(\ref{eq:gaussianHeis}), with $Q_c=50$,
  $s=10$, $k=\pi/2$, with uniform coupling.  Note how the evolution is
  dispersion-free.}
\label{fig:twoHeisenbergs}
\end{figure}
}

\newcommand{\placeFigFive}{\begin{figure}[t!]
\centering
\includegraphics[height=6.7cm]{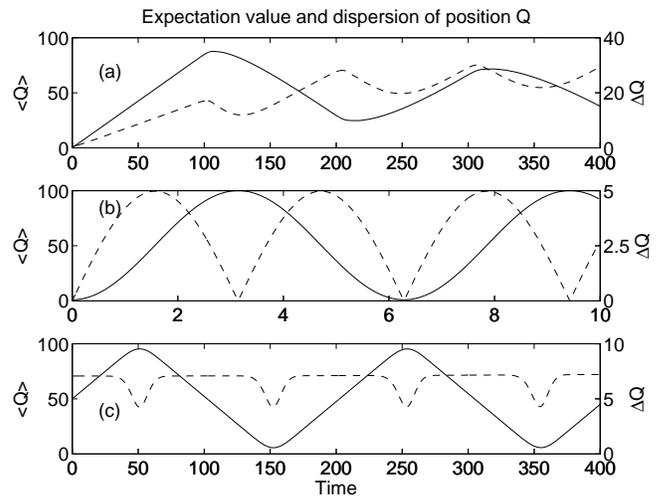}\\
\caption{Evolution of a Heisenberg spin chain Hamiltonian
  [Eq.~(\ref{eq:Heis1up})], with $J=1$.  The solid lines display the
  expectation value of the position $\langle Q\rangle$, the dashed
  lines show the dispersion $\Delta Q$.  (a) Evolution of $|1\rangle$
  under uniform coupling, $\Delta Q$ increases with time, this
  indicates increasing dispersion as time progresses.  (b) Evolution
  of $|1\rangle$ under parabolic coupling, here $\Delta Q$ does not
  increase with time, as such this indicates dispersion-free
  evolution. (c) Evolution of a Gaussian pulse as in
  Eq.~(\ref{eq:gaussianHeis}), with $Q_c=50$, $s=10$, $k=\pi/2$, with
  uniform coupling.  As in (b), there is no overall increase in
  dispersion, as such this pulse is also dispersion-free.}
\label{fig:Q}
\end{figure}
}

\newcommand{\placeFigSix}{\begin{figure*}[th!]
\centering
\includegraphics[width=\wide]{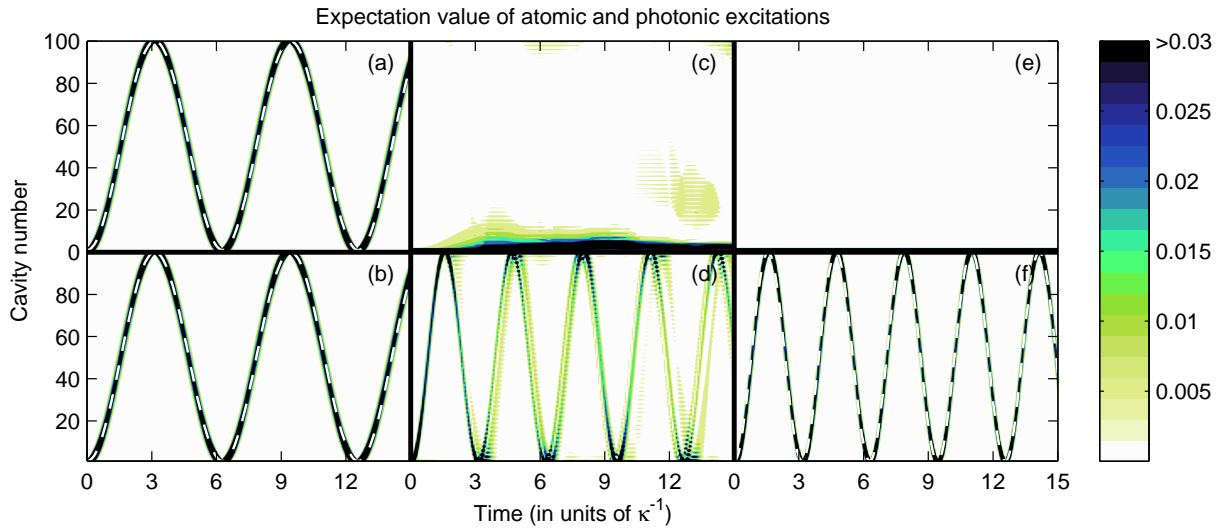}
\caption{ (Color online) Space-time diagrams for evolution of an
  excitation ($|1\rangle\otimes(|g,1\rangle+|e,0\rangle)/\sqrt{2}$)
  along a chain of 100 JC cavities with parabolic inter-cavity
  coupling profile.  The upper plots show the population of the atomic
  components whereas the lower plots are the photonic components.  The
  ratio of cavity-cavity coupling to atom-photon coupling varies from
  left to right, with (a) and (b) $\kappa/\beta=10^{-4}$, (c) and (d)
  $\kappa/\beta=10^0$, (e) and (f) $\kappa/\beta=10^3$.  In plots (a),
  (b) and (f), the white dashed lines represent $\langle Q(t)\rangle$,
  as given in Eq.~(\ref{eq:Qev}).  The effect of the parabolic
  coupling is to constrain the pulse, resulting in well defined and
  reversible dispersion.}
\label{fig:bigParabolicBottom}
\end{figure*}
}

\newcommand{\placeFigSeven}{\begin{figure*}[th!]
\centering
\includegraphics[width=\wide]{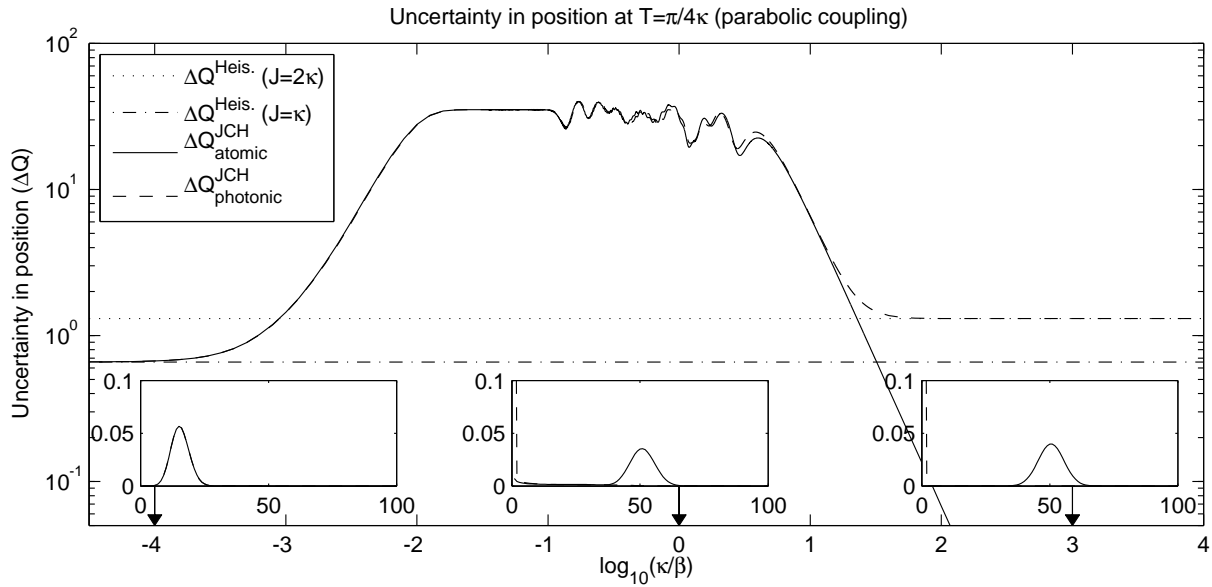}
\caption{ Dispersion of the wave packet, $\Delta Q$, at a fixed point
  in time for a JCH system with parabolic coupling.  The solid
  (dashed) line shows $\Delta Q_{\rm atomic}^{\rm JCH}$ ($\Delta
  Q_{\rm photonic}^{\rm JCH}$).  The dotted (dot-dashed) line shows
  $\Delta Q^{\rm Heis}$ with $J=2\kappa$ ($J=\kappa$).  The parameter
  ranges are identical to Fig.~(\ref{fig:bigUniformTop}), although the
  range of the vertical axis is different.  The insets show the pulse
  profile at three values of $\kappa/\beta$.  The solid (dashed) line
  represents the photonic (atomic) profile. The result is that, while
  the dispersion is considerably larger than the uniform coupling
  case, in this case the pulse is more `well behaved'.  More
  precisely, the pulse has a Gaussian profile away from the boundaries
  and its evolution is completely reversible. In the extreme limits
  for $\kappa$, we have again perfect agreement with that expected
  from the appropriate spin chain model.  }
\label{fig:bigParabolicTop}
\end{figure*}
}

\newcommand{\placeFigEight}{\begin{figure*}[th!]
\centering
\includegraphics[width=\wide]{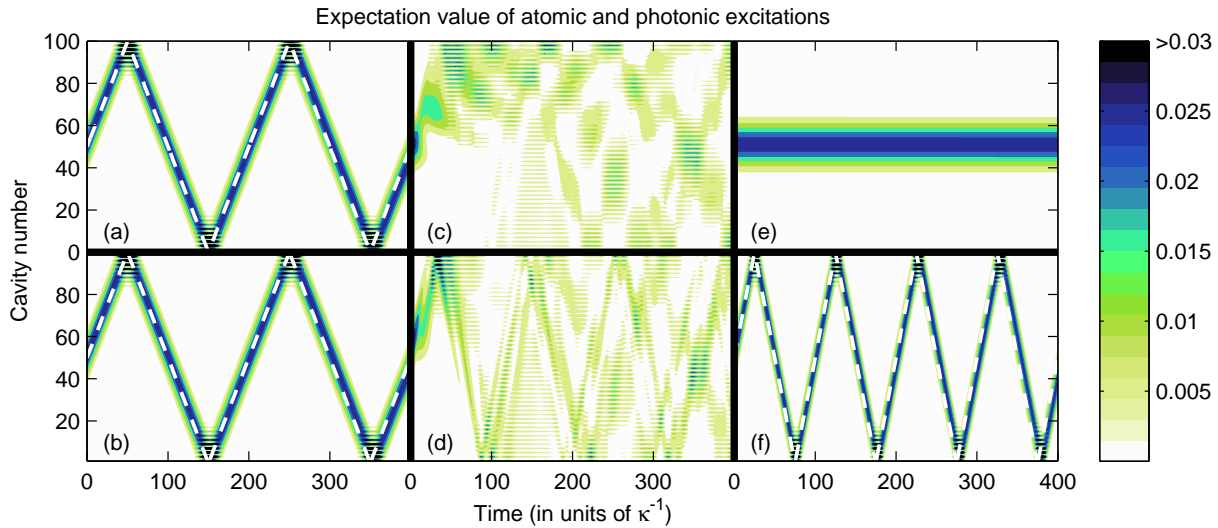}
\caption{ (Color online) Space-time diagrams for evolution of an
  Gaussian pulse along a chain of 100 JC cavities with a uniform
  inter-cavity coupling profile.  The upper plots show the population
  of the atomic components whereas the lower plots are the photonic
  components.  The ratio of cavity-cavity coupling to atom-photon
  coupling varies from left to right, with (a) and (b)
  $\kappa/\beta=10^{-2}$, (c) and (d) $\kappa/\beta=10^0$, (e) and (f)
  $\kappa/\beta=10^3$.  In plots (a), (b) and (f), the white dashed
  lines represent $\mathcal{Q}^{\rm Gaussian}_{\wedge}$, as given in
  Eq.~(\ref{eq:triangleGaussian}).  The effect of the Gaussian pulse
  is to have chosen a narrow momentum distribution, which allows the
  excitation to propagate freely in the system.}
\label{fig:bigGaussianBottom}
\end{figure*}
}

\newcommand{\placeFigNine}{\begin{figure*}[th!]
\centering
\includegraphics[width=\wide]{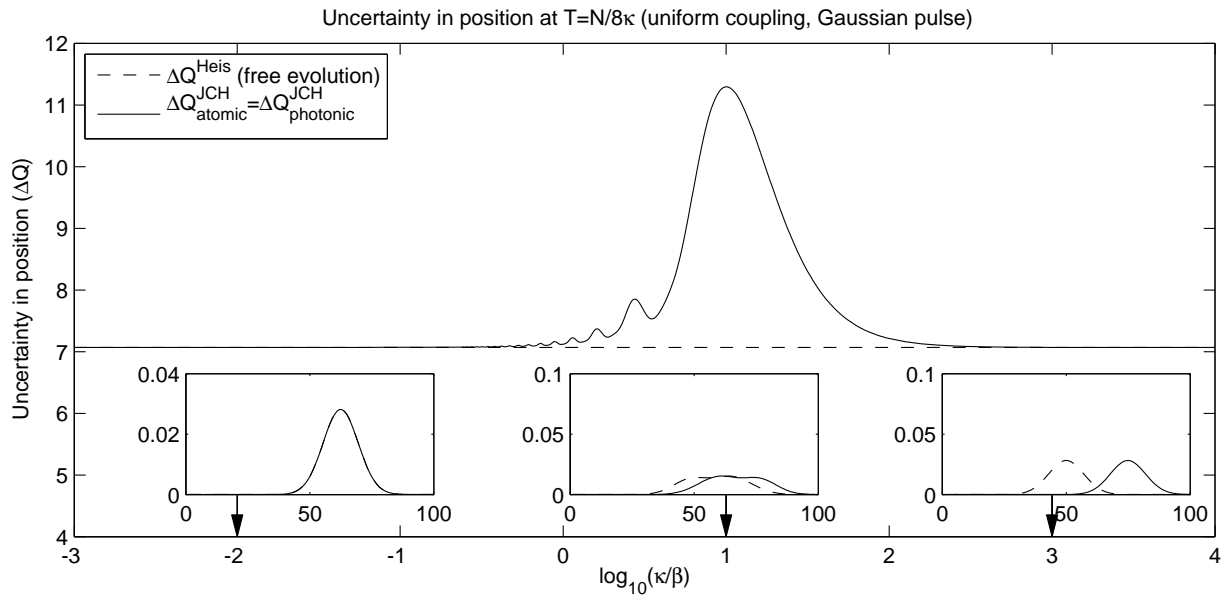}
\caption{ Dispersion of a Gaussian wave packet, $\Delta Q$, at a fixed
  point in time, as a function of $\kappa/\beta$, at $\Delta=0$.  The
  dispersion is measured at time $T=N/8\kappa$, at which point the
  packet is `freely' evolving along the chain.  The initial state of
  the Gaussian is given by Eq.~(\ref{eq:gaussianJCH}), with center of
  the pulse at $Q_c=N/2$, width of the pulse $s=N/10$, wave number
  $k=\pi/2$, and number of cavities $N=100$.  The width of the pulse
  is small enough such that the boundaries do not have an effect on
  $\Delta Q$ for the value of $T=N/8\kappa$ in the limits
  $\kappa/\beta \ll 1$ and $\kappa/\beta \gg 1$.  At the left of the
  plot, $\kappa/\beta \ll 1$, and the JCH chain mimics two identical
  Heisenberg spin chains, an example is shown of this localized
  behavior in Fig.~(\ref{fig:bigGaussianBottom})(a) and (b), where
  $\kappa/\beta=10^{-2}$.  The corresponding spatial profile of the
  pulse is shown in the left inset, (the solid line shows the photonic
  profile and the dashed line shows the atomic profile, in this case
  they are coincident).  At the right of the plot $\kappa/\beta\gg1$,
  and the photonic mode of the JCH chain mimics a Heisenberg spin
  chain, while the atomic mode does not propagate at all.  An example
  of this type of localized behavior is shown in
  Fig.~(\ref{fig:bigGaussianBottom})(e) and (f), where
  $\kappa/\beta=10^3$, the corresponding profile of the pulse at
  $T=N/8\kappa$ is shown in the right inset.  In the middle of the
  plot a travelling excitation shows a large amount of dispersion, an
  example of this delocalized behavior is shown in
  Fig.~(\ref{fig:bigGaussianBottom})(c) and (d), where
  $\kappa/\beta=10$, and the corresponding profile at $T=N/8\kappa$ is
  shown in the middle inset.  The horizontal line shows the dispersion
  $\Delta Q^{\rm Heis}$ of a freely evolving pulse [this value does
    not change, provided the pulse is not interacting with the
    boundary, as is evidenced by Fig.~(\ref{fig:Q})(c)].}
\label{fig:bigGaussianTop}
\end{figure*}
}

\newcommand{\placeFigTen}{\begin{figure}[t!]
\centering
\includegraphics[width=\wid]{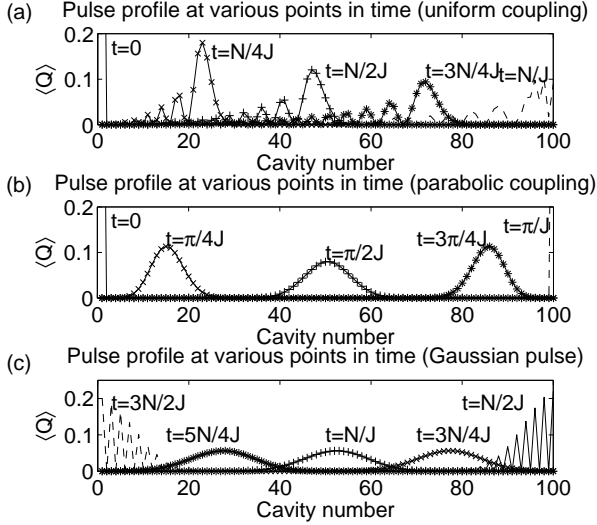}\\
\caption{The pulse profile for various points in time as a function of
  cavity number, for evolution of a Heisenberg spin chain.  (a)
  Evolution of $|1\rangle$ with uniform coupling between spins.  (b)
  Evolution of $|1\rangle$ with parabolic coupling between spins. (c)
  Evolution of a Gaussian pulse, as given in
  Eq.~(\ref{eq:gaussianHeis}) with uniform coupling between spins.  In
  the first uniform case, the profile is initially given by a
  Kronecker delta function, but later spreads into a function with one
  primary peak and a number of smaller trailing peaks.  In the
  parabolic case, at each end of the chain, the pulse is given by a
  Kronecker delta function, while in the middle of the chain the pulse
  approximates a Gaussian.  In the uniform coupling case with an
  initial Gaussian pulse, it evolves along the chain with fixed
  profile, and at the ends of the chain interferes and changes
  direction.}
\label{fig:pulseProfiles}
\end{figure}
}

\newcommand{\placeFigEleven}{\begin{figure}[h!]  \centering
    \includegraphics[width=7.3cm]{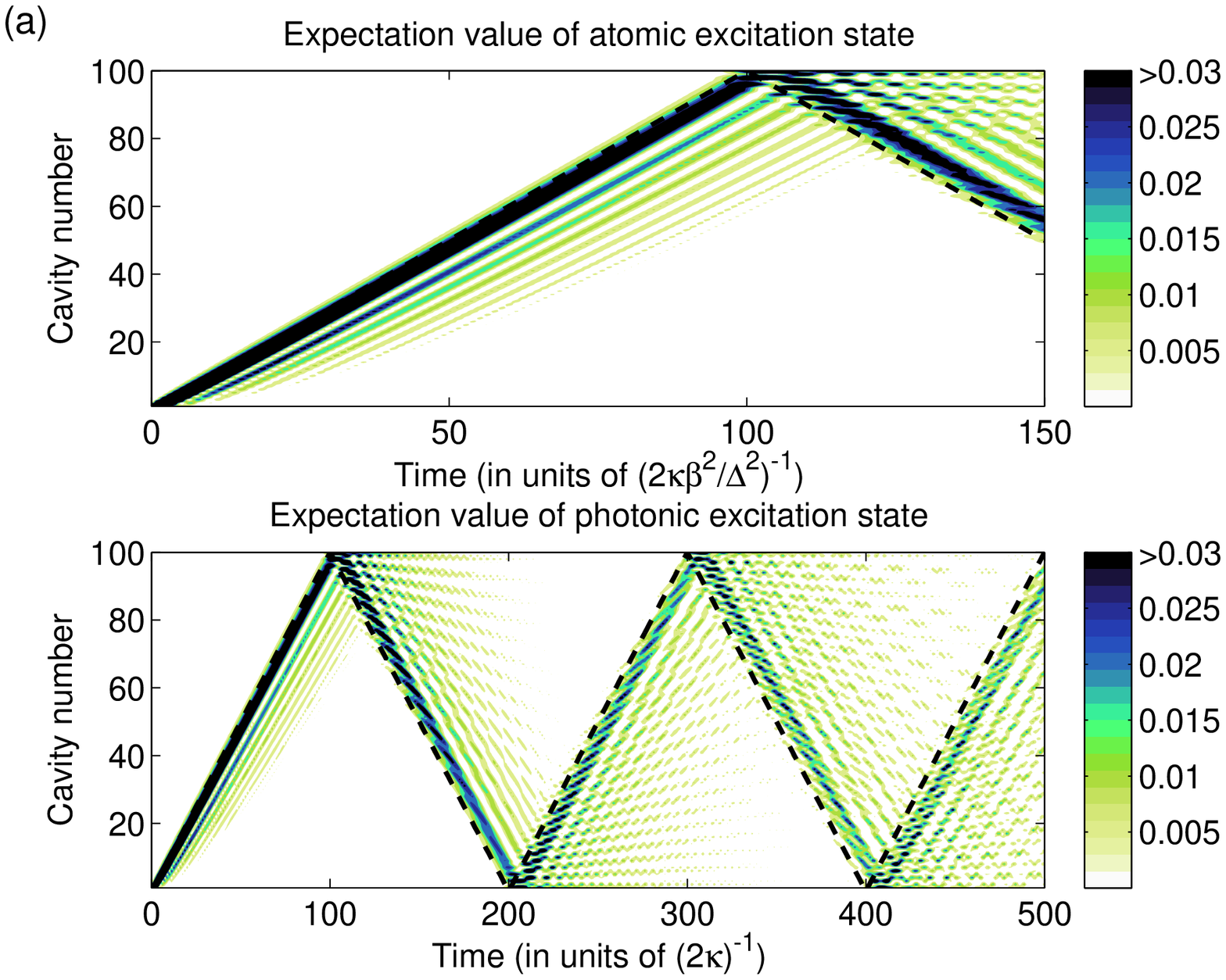}\\\includegraphics[width=7.3cm]{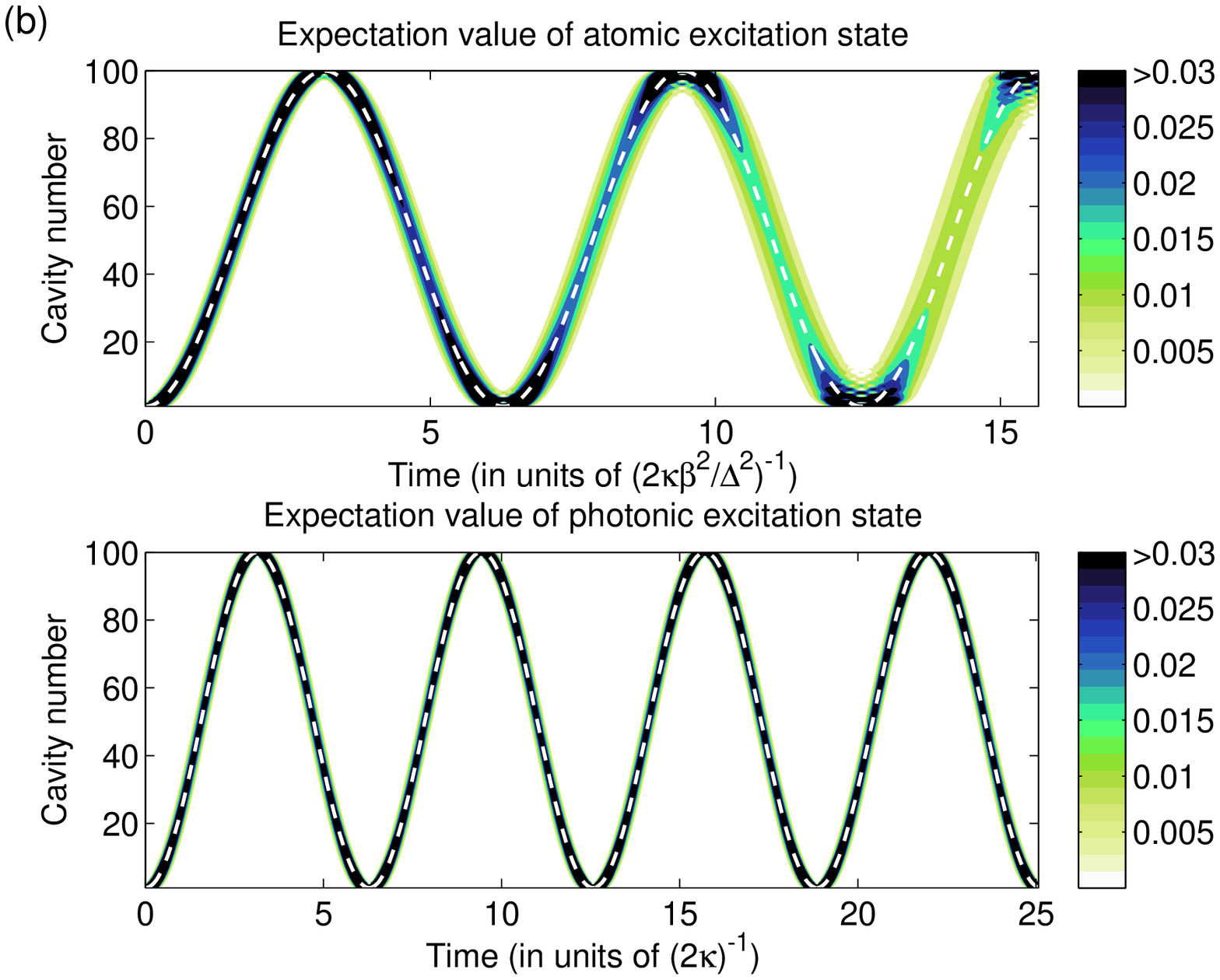}\\\includegraphics[width=7.3cm]{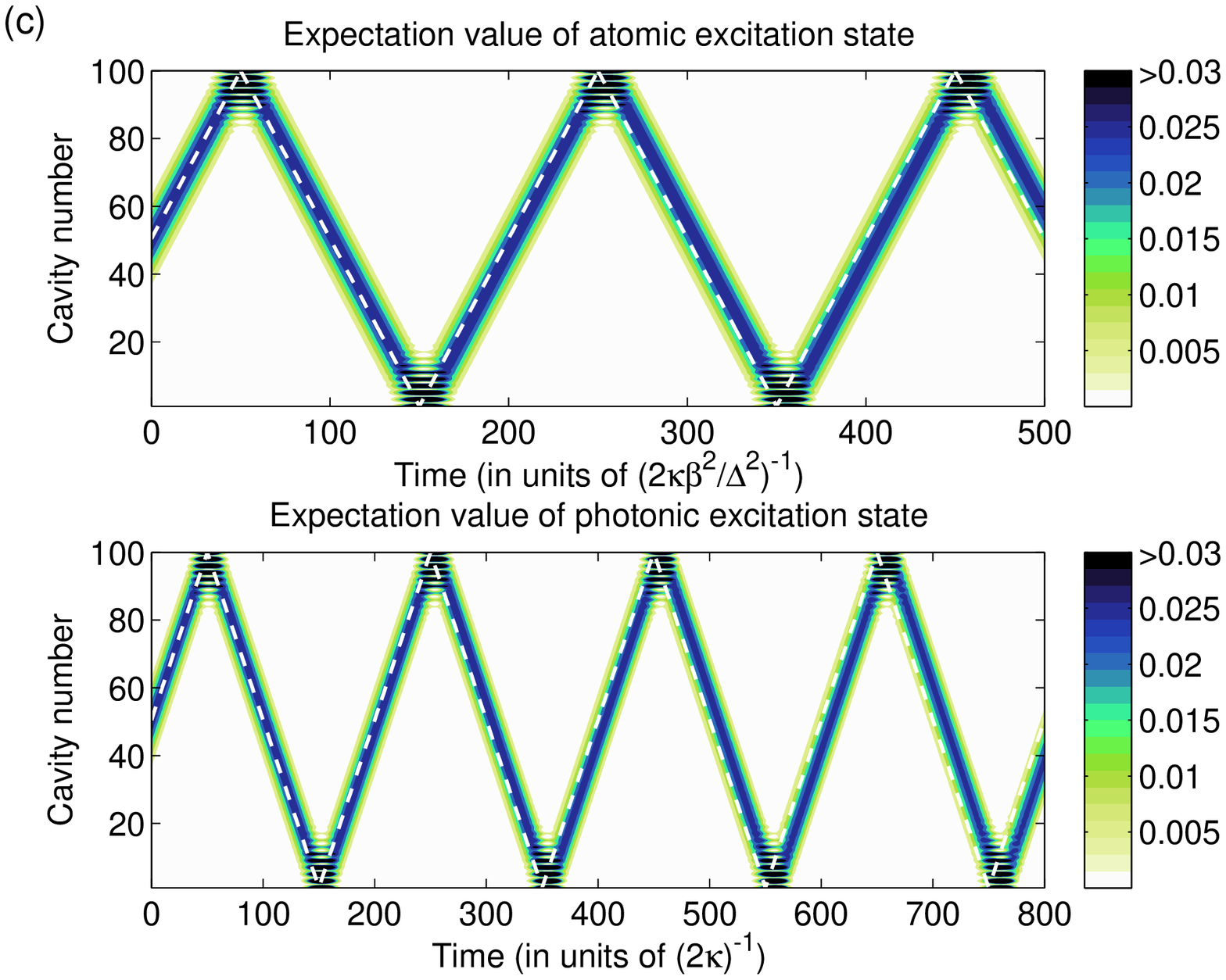}
\caption{(Color online) These plots show the evolution of the state
  $|1\rangle\otimes(|e,0\rangle + |g,1\rangle)$ (a) and (b) and a
  Gaussian pulse (c) in a one-dimensional JCH system consisting of 100
  cavities, with $\kappa=\beta$ and $\Delta/\beta = 10^3$.  Note the
  time scales are different in each case.  (a) Evolution in a uniform
  chain, where the dashed lines are a triangle wave giving approximate
  evolution of the wavefront, as given in Eq.~(\ref{eq:triangle}). (b)
  Evolution in a parabolically coupled chain, where the dashed lines
  are as given in Eq.~(\ref{eq:Qev}). (c) Evolution in a uniformly
  coupled chain, with an initial Gaussian pulse, as given in
  Eq.~(\ref{eq:gaussianJCH}). The dashed lines show the expectation
  value of position, as given in Eq.~(\ref{eq:triangleGaussian}).}
\label{fig:differentSpeeds}
\end{figure}
}

\title{Time evolution of the one-dimensional Jaynes-Cummings-Hubbard Hamiltonian}
\author{M.I. Makin}
\affiliation{Center for Quantum Computer Technology, School of Physics, The University of Melbourne, Victoria 3010, Australia}

\author{Jared H. Cole}
\affiliation{Institut f\"ur Theoretische Festk\"orperphysik and DFG-Center for Functional Nanostructures (CFN), Universit\"at Karlsruhe, 76128, Germany}

\author{Charles D. Hill}
\affiliation{Center for Quantum Computer Technology, School of Physics, The University of Melbourne, Victoria 3010, Australia}

\author{Andrew D. Greentree}
\affiliation{Center for Quantum Computer Technology, School of Physics, The University of Melbourne, Victoria 3010, Australia}

\author{Lloyd C. L. Hollenberg}
\affiliation{Center for Quantum Computer Technology, School of Physics, The University of Melbourne, Victoria 3010, Australia}

\begin{abstract}
The Jaynes-Cummings-Hubbard (JCH) system describes a network of
single-mode photonic cavities connected via evanescent coupling.  Each
cavity contains a single two level system which can be tuned in
resonance with the cavity.  Here we explore the behavior of single
excitations (where an excitation can be either photonic or atomic) in
the linear JCH system, which describes a coupled cavity waveguide.  We
use direct, analytic diagonalization of the Hamiltonian to study cases
where inter-cavity coupling is either uniform or varies parabolically
along the chain. Both excitations located in a single cavity, as well
as one excitation as a Gaussian pulse spread over many cavities, are
investigated as initial states. We predict unusual behavior of this
system in the time domain, including slower than expected propagation
of the excitation, and also splitting of the excitation into two
distinct pulses, which travel at distinct speeds.  In certain limits,
we show that the JCH system mimics two Heisenberg spin chains.
\end{abstract}

\maketitle

\section{Introduction}

The realization of condensed matter models using non-traditional
controllable systems has been an active topic of recent research.
This is best exemplified by the work on cold atom quantum simulators
\cite{ref:bloch, ref:balili}.  However the possibility of quantum
optical simulators for condensed matter models has emerged.  While
finding links and analogies between different subfields of physics is
interesting in its own right, of particular interest are effects that
are difficult or impossible to observe in more conventional physical
systems.

There are a number of recent proposals \cite{ref:hartmannNP, ref:GTCH,
  ref:angelakis07, ref:pointylobes, ref:BoseAngelakisBurgath,
  ref:irish, ref:ogden, ref:koch} for constructing quantum optical
condensed matter simulators.  These use coupled cavity structures
where confined photons are induced to interact via their coupling to
embedded two-state systems.  Possible two-state systems include color
centers \cite{ref:GTCH}, quantum dots
\cite{ref:na2yamamoto,ref:grochol}, superconducting strip-line
resonators \cite{ref:wallraff, ref:jisunxieliu, ref:gerace}, or
coupled Rubidium microcavities \cite{ref:trupke}.

In such systems, it should be possible to observe many-body effects
such as quantum phase transitions where the particles of interest are
photons (rather than electrons) and therefore can be readily injected,
confined and observed individually.  While many-body effects in the
thermodynamic limit require a large number of cavities and/or photons
(often solved in the mean field limit), it is quite clear that early
experiments will be limited in both overall structure size and system
controllability.  It is therefore of interest to determine what (if
any) effects can be observed in the few cavity/few excitation limit
(where an excitation can either be photonic or atomic)
\cite{ref:irish, ref:shenfanPRL, ref:shenfanOL, ref:phasetransitions}.
Importantly, the complex interactions between atomic and photonic
components produce a variety of effects.  Solitonic behavior has been
predicted in the 1D coupled cavity wave guide realizations of the
Dicke model \cite{ref:kimsoliton} and the XXZ model in the presence of
a tilted magnetic field \cite{ref:sunsoliton}.  If one is considering,
for example, the type of localized/delocalized behavior typically
found in strongly interacting systems, it is of vital importance not
to be distracted by the background effects which come from the
delocalized behaviour of photons themselves.  For this reason, a
detailed understanding of the single-particle dynamics is paramount.

Here we investigate the time evolution of a one dimensional coupled
cavity waveguide \cite{ref:altug} described by the
Jaynes-Cummings-Hubbard (JCH) model \cite{ref:GTCH}. In the single
excitation subspace we diagonalize the JCH Hamiltonian exactly and
consider the dependence of this system on three parameters:
atom-cavity detuning, atom-photon coupling and coupling between
cavities.  We then investigate limits in which localized and
delocalized behaviour can be seen.  In this paper, we refer to the
propagation of photonic or atomic excitations as the photonic and
atomic components, or modes, respectively (in the atomic case, the
atoms remain stationary whilst the excitation passes through).  In
particular we focus on three limits.  Firstly, the limit where the
atom-cavity detuning is zero and the coupling strength between
cavities is much less than the photonic cavity-atomic coupling.  In
this limit, the propagation dynamics of the atomic and photonic
components are identical; we find they propagate as a pulse travelling
back and forth along the line of cavities.  Secondly, we study the
limit where photon hopping dominates all other parameters of the
system.  In this limit, the atomic component does not move, while the
photonic component propagates.  Thirdly, we study the limit where the
atom-cavity detuning is much larger than all other energy scales of
the system.  In this limit the atomic and photonic modes travel at two
different speeds.  In these three limits we compare the behaviour with
that of two uncoupled Heisenberg spin chains
\cite{ref:angelakis07,ref:mattis,ref:sachdev}, in the one-excitation
case.

We discuss the JCH model and the uniform coupling case in
Sec.~\ref{sec:uniform}. We then continue with dispersion-free pulses
using parabolic coupling, Sec.~\ref{sec:parabolic}, and initial
Gaussian pulses, Sec.~\ref{sec:gaussian}.  In
Sec.~\ref{sec:largedetuning} we discuss the limit of large atom-cavity
detuning, in this limit the atomic and photonic modes travel at two
distinct speeds.

\placeFigOne

\section{Uniform coupling}
\label{sec:uniform}

The JCH Hamiltonian ($\Ha^{\rm JCH}$) describes a system of $N$
cavities linked via photon hopping under the tight-binding
approximation \cite{ref:GTCH, ref:angelakis07,
  ref:phasetransitions}. A possible arrangement and schematic showing
the couplings is shown in Fig.~(\ref{fig:1DJCH}).  The Hamiltonian is

\begin{equation}
\label{eq:mainham}
\Ha^{\rm JCH} = \sum_{i=1}^N \Ha^{\rm JC}_i -  \kappa \sum_{i,j=1}^{N}A_{ij} a^{\dagger}_i a_j,
\end{equation}
where $\kappa$ is the inter-cavity coupling, and $A$ is the adjacency
matrix, which is defined according to the geometry and boundary
conditions of the system. Using the geometry as implied by
Fig.~(\ref{fig:1DJCH})(a), that is, one spatial dimension with hard
wall boundary conditions, $A$ is given by

\begin{equation}
\label{eq:adjacency}
A_{ij} = \left\{ \begin{array}{lll}
1, & {\rm if }& |i-j|=1\\
0, & & \mbox{otherwise}
\end{array}\right..
\end{equation}

  The Jaynes-Cummings Hamiltonian $\Ha^{\rm
  JC}$ \cite{ref:jaynescummings} describes a single atom-cavity
system.  It can be written 

\begin{equation}
\label{eq:JCham}
\Ha^{\rm JC}_i = \epsilon_i \sigma^+_i \sigma^-_i + \omega_i a^{\dagger}_i a_i + \beta_i (\sigma^+_i a_i + \sigma^-_i a^{\dagger}_i),
\end{equation}
where $a^{\dagger}_i$ ($a_i$) and $\sigma^+_i$ ($\sigma^-_i$) are the
photonic and atomic raising (lowering) operators respectively, and
$\hbar=1$.  In cavity $i$, the energy of the atom is given by
$\epsilon_i$, the cavity resonance by $\omega_i$, and the cavity-atom
coupling by $\beta_i$.

The single cavity {\it bare basis} consists of states of the form
$|s,n\rangle$, where $s\in\{g,e\}$ represents the two-level atom in
the ground or excited state, and $n$ is a non-negative integer,
representing the number of photons in the cavity.  The $N$ cavity
bare basis simply consists of the tensor product of $N$ such single
cavity bare bases.

The single cavity {\it dressed basis} consists of the state
$|g,0\rangle$ (this state is a member of both the bare and dressed
bases) and states of the form $|\pm,n\rangle$, where $n$ is the number
of excitations (photonic or atomic) in the cavity.  They are energy
eigenstates of the Jaynes-Cummings Hamiltonian, and are related to the
single cavity bare basis states by

\begin{equation}
\begin{split}
 |\pm,n\rangle = \frac{\beta\sqrt{n} |g,n\rangle + [-(\Delta/2)\pm \chi(n)]|e,n-1\rangle}{\sqrt{2\chi^2(n)\mp \chi(n)\Delta}}\\
 \quad \forall n\geq 1,
\label{eq:dressed}
\end{split}
\end{equation}
where $\Delta=\omega-\epsilon$ is the detuning, and we have used the
generalized Rabi frequency

\begin{equation}
\label{eq:rabi}
\chi(n) = \sqrt{n\beta^2 + \Delta^2/4}\quad \forall n\geq 1.
\end{equation}

\placeFigTwo

Consider a restricted basis of the full $N$ cavity bare basis,
consisting only of basis states that contain one excitation (atomic or
photonic) in the whole system: the one excitation subspace.  The total
number of excitations in the JCH system is conserved, hence this is a
restriction, not an approximation.  We consider the case where all
cavities are equal, hence $\epsilon_i=\epsilon, \omega_i=\omega,
\beta_i=\beta$ for all $i=1,\ldots,N$.  We write the restricted one
excitation basis as $|Q\rangle\otimes|g,1\rangle$
($|Q\rangle\otimes|e,0\rangle$), where $Q\in \{1,\ldots,N\}$, to mean
a photonic excitation $|g,1\rangle$ (atomic excitation $|e,0\rangle$)
at cavity number, $Q$, and $|g,0\rangle$ at every other cavity
position.  The states $|Q\rangle$ form a valid Hilbert space.  In this
restricted basis and under these conditions, the Hamiltonian
(\ref{eq:mainham}) is greatly simplified, and can be represented

\begin{equation}
\label{eq:JCH1exc}
\mathcal{H}^{\rm JCH}_{\rm 1exc} = \frac{\Delta}{2} I_N\otimes Z + \beta I_N\otimes X - \kappa A \otimes\frac{I_2+Z}{2},
\end{equation}
where $I_m$ is the $m\times m$ identity matrix, $X$ and $Z$ are the
usual Pauli matrices acting on the atom-photon cavity subspace. The
operator $A$ in matrix form is given by the adjacency matrix of the
connectivity graph, for example that given in
Eq.~(\ref{eq:adjacency}).  In this form, the first and second terms
affect only the atomic/photonic modes locally, but do not move the
excitation to any other cavity.  The first term describes the detuning
and the second term the coupling between photonic and atomic
excitation modes.  The third term describes photonic coupling between
cavities.

By diagonalizing $A$ the whole JCH Hamiltonian restricted to one
excitation can be diagonalized.  We now solve for the 1D chain with
hard wall boundary conditions.  The eigenvectors of the $A$ matrix
given in Eq.~(\ref{eq:adjacency}) are \cite{ref:christandletal}

\begin{equation}
|k\rangle =   \frac{\sqrt{2}(-1)^k\sin\left(\frac{N k\pi}{N+1}\right)}{\sqrt{N+1}\sin\left(\frac{k\pi}{N+1}\right)}\sum_{Q=1}^N \sin\left(\frac{Qk\pi}{N+1}\right) |Q\rangle,
\end{equation}
where $k=1,\ldots,N$.

We now wish to obtain the energy eigenstates and energy eigenvalues
for the entire one excitation subspace. For a linear chain of cavities
the Hamiltonian $\Ha^{\rm JCH}_{\rm 1exc}$ can be expressed in the
basis $\{|k\rangle\otimes|g,1\rangle$,
$|k\rangle\otimes|e,0\rangle\}$, $k=1,\ldots,N$ as a block diagonal
matrix, in which the $k$th block appears as

\begin{equation}
\mathcal{H}^{\rm JCH}_{\rm 1exc}(k) =
\left(\begin{array}{cc}
\Delta/2 + 2\kappa\cos[ k\pi/(N+1)] & \beta\\
\beta & -\Delta/2
\end{array}
\right).
\end{equation}

\placeFigThree

The eigenvalues of the full Hamiltonian, from Eq.~(\ref{eq:JCH1exc}), are

\begin{equation}
E_{\pm}^k = \kappa \cos \left(\frac{k\pi}{N+1}\right) \pm \sqrt{\left[\frac{\Delta}{2}+\kappa\cos\left(\frac{k\pi}{N+1}\right)\right]^2+\beta^2},
\end{equation}
where the second term of this appears very similar to the Rabi
frequency $\chi(1)$, Eq.~(\ref{eq:rabi}), with the detuning $\Delta$
shifted by the cosine term.  These eigenvalues have corresponding
eigenvectors

\begin{equation}
\label{eq:evecs}
|\pm,k\rangle =  \left[ \frac{ (\Delta+2E_{\pm}^k) |e,0\rangle + 2\beta |g,1\rangle}{ \sqrt{(\Delta+2E_{\pm}^k)^2+4\beta^2}}\right]\otimes|k\rangle .
\end{equation}

This is an exact diagonalization of the linearly-coupled JCH system in
the one excitation subspace.  Hence, it is possible to determine
analytically, for arbitrary $N$, the time-evolution of an arbitrary
initial state.  In the case with two cavities, i.e.~$N=2$, and in
appropriate limits to each case, this evolution recovers equations
(22), (28) and (29) of Ogden et al.~\cite{ref:ogden}.  Specifically,
we are interested in plotting the expectation value of both the
photonic and atomic excitations in cavity $j$, by examining the number
operators $a_j^{\dagger}a_j$ and $\sigma^+_j\sigma^-_j$.

As a demonstration of the dynamics of the system when on resonance
(i.e.~$\Delta=0$), we consider evolution of an excitation initially located
in the first cavity in a line of 100 cavities, in an equal
superposition of atomic and photonic modes, i.e.

\begin{equation}
|\psi(t=0)\rangle = |1\rangle\otimes(|g,1\rangle +|e,0\rangle)/\sqrt{2},
\label{eq:equalsuper}
\end{equation}
which corresponds to the single cavity energy eigenstate $|+,1\rangle$
from Eq.~(\ref{eq:dressed}) in the case $\Delta=0$.  As such, there
will be no evolution within a cavity, but only evolution between
cavities.  We solve the evolution exactly, using the usual
Schr\"odinger equation, $i\partial_t |\psi\rangle = H |\psi\rangle$.
This state also constitutes one of the simplest states to realize
experimentally, as the JC resonance can be driven directly by a
transverse field.  The dispersion and other effects discussed in this
paper stem directly from the strong atom-photon interaction via the JC
Hamiltonian.  If we were to consider a system with vanishing coupling
to the atoms ($\beta=0$), we regain the conventional photon
propagation results \cite{ref:shenfanPRL, ref:shenfanOL} in which a
Gaussian wave propagates smoothly.

\placeFigFour

We are ultimately interested in the behavior of the system when
various energies dominate, such as the atom-photon coupling or the
cavity-cavity coupling.  For this reason, we will now consider the
system in several different limits.  When $\kappa/\beta\ll 1$ and
$\Delta=0$, the atomic and photonic modes have identical propagation
dynamics. The case $\kappa/\beta=10^{-3}$ is shown in
Fig.~(\ref{fig:bigUniformBottom})(a) and (b).  We may understand the
equal propagation because the atom-photon coupling $\beta$ is much
stronger than the cavity-cavity coupling $\kappa$, hence the
excitation is free to form the single cavity eigenstate between
inter-cavity hops. By shifting to the interaction picture
\cite{ref:mikeandike}, that is $|\chi\rangle = e^{i \beta I\otimes X
  t} |\psi\rangle$, the Hamiltonian Eq.~(\ref{eq:JCH1exc}) becomes

\begin{equation}
\label{eq:deltaZero}
H = -\frac{\kappa}{2} A \otimes I,
\end{equation}
where the fast rotating terms have been ignored.  It is useful at this
point to compare with the Hamiltonian that describes the well-known
Heisenberg spin chain \cite{ref:fisher1989,ref:stanley}.

\begin{eqnarray}
 H^{\rm Heis} \!\!\!&=&\!\!\! -J \sum_{n=1}^N \mathbf{S}_n . \mathbf{S}_{n+1} \\
&=& \!\!\!-J \sum_{n=1}^N \left[ \frac{1}{2} ( S_n^+ S_{n+1}^- + S_n^-
    S_{n+1}^+) + S_n^z S_{n+1}^z\right]\nonumber.
\end{eqnarray}

The single site basis of this system consists of spins pointing up and
down along the direction of the $z$ axis,
$\{|\!\!\uparrow\rangle,|\!\!\downarrow\rangle\}$. The $N$ site basis
comprises of a tensor product of $N$ such bases. This Hamiltonian also
conserves the total spin in the $z$ direction. So, if we limit the $N$
site basis to having only one $|\!\!\uparrow\rangle$, and the rest
$|\!\!\downarrow\rangle$, then the Hamiltonian can be represented in
this restricted subspace as (ignoring contributions from the $S^z$
term, which is largely a phase factor apart from a minor shift at the
ends of the chain, this is inconsequential for long chains)

\begin{equation}
\label{eq:Heis1up}
H^{\rm Heis} = -\frac{J}{2} A,
\end{equation}
where the basis vectors are now
$|1\rangle,|2\rangle,\ldots,|N\rangle$, and $|Q\rangle$ represents
$|\!\!\uparrow\rangle$ at site $Q$ and $|\!\!\downarrow\rangle$ at
every other site, and $A$ is the adjacency matrix, for example that
which was introduced in Eq.~(\ref{eq:adjacency}).  The initial state
$|1\rangle$ will evolve along the chain (due to the hard wall boundary
conditions) with an approximate speed $J$ (the rate at which the front
of the excitation wave travels across the chain), according the
triangle wave

\begin{equation}
\mathcal{Q}_{\wedge} = \frac{N-1}{\pi} \arcsin\left\{\sin\left[\pi\left(\frac{J t}{N} -
  \frac{1}{2}\right)\right]\right\} + \frac{N+1}{2}.
\label{eq:triangle}
\end{equation}
We find that the speed $J$ as described above is given by the
derivative of Eq.~(\ref{eq:triangle}), i.e.~
$J=|\partial_t\mathcal{Q}|$ (where defined).

\placeFigFive
By comparing Eq.~(\ref{eq:Heis1up}) with Eq.~(\ref{eq:deltaZero}), we
see that the JCH Hamiltonian in the regime $\Delta=0$,
$\kappa\ll\beta$, mimics two Heisenberg spin chains, in each of the
photonic and the atomic states.  In both cases, $J=\kappa$ (the
approximate excitation speed).

Next we consider an alternative limit, where the coupling between
cavities dominates the evolution, $\kappa/\beta\gg 1$.  In this
regime, the atomic mode does not propagate at all, while the photonic
mode propagates at twice the speed of the previous case.  This is
because the atom-photon coupling is effectively zero compared to the
much faster inter-cavity coupling rate, freezing the atomic
excitation. The case $\kappa/\beta = 10^3$ is shown in
Fig.~(\ref{fig:bigUniformBottom})(e) and (f).  In this limit, the
Hamiltonian trivially reduces to

\begin{equation}
H = -\kappa A \otimes \frac{I_2 + Z}{2},
\end{equation}
as such, the photonic excitation travels with speed $2\kappa$, and the
atomic excitation does not move at all.  We can think of this limit as that of a pure photon gas, albeit with only one photon, which is equivalent to a single-excitation spin chain.

\placeFigSix

When $\Delta=0$, and $\kappa$ and $\beta$ are of the same order of
magnitude, the evolution of the state Eq.~(\ref{eq:equalsuper}) now
experiences a large amount of dispersion.  This is because the
evolution no longer approximates two Heisenberg spin chains (as it
does in the $\kappa/\beta\ll1$ and $\kappa/\beta\gg1$ limits), but
rather the full JCH nature of the evolution is expressed.  As the
atom-cavity coupling $\beta$ and the cavity-cavity coupling $\kappa$
are of the same order, there is no opportunity for the excitation to
form any single cavity eigenstates, instead the excitation is
relatively free to roam between photonic and atomic modes, as well as
between cavities.  The example $\kappa/\beta=10$ is shown in
Fig.~(\ref{fig:bigUniformBottom})(c) and (d).

To more clearly see this interplay between JC and photon dominated
regimes, we consider the dispersion of the wave packet as it travels
along the chain.  Fig.~(\ref{fig:bigUniformTop}) shows how the
dispersion of a pulse changes with $\kappa/\beta$.  We define the
position operators

\begin{equation}
\begin{split}
Q_{\rm photonic} &\equiv {\rm diag}(1,0,2,0,\ldots,N,0)\\
Q_{\rm atomic}   &\equiv {\rm diag}(0,1,0,2,\ldots,0,N)
\label{eq:Qphotonicatomic}
\end{split}
\end{equation}
in the basis $\{|Q\rangle\otimes|g,1\rangle,
|Q\rangle\otimes|e,0\rangle\}$, $Q=1,\ldots,N$.  The definitions of
Eq.~(\ref{eq:Qphotonicatomic}) assume that when calculating
expectation values for photonic (atomic) position, the photonic
(atomic) states are selected from the wave function and normalized as
a single vector of length $N$ \footnote{Imagine measuring two
  qualities on each of an infinite set of identically prepared 1D JCH
  systems.  The first quality is the atomic/photonic nature of the
  excitation.  The second quality is position.  Averaging the position
  reading of all photonic (atomic) systems gives the expectation value
  of the photonic (atomic) position, as given in
  Eq.~(\ref{eq:Qphotonicatomic}).}.

The dispersion for photonic and atomic modes is given by the standard
deviation of the position operator $\Delta Q$

\begin{equation}
(\Delta Q_{\rm type})^2 = \langle Q_{\rm type}^2\rangle - \langle
  Q_{\rm type}\rangle^2.
\end{equation}

Looking at Fig.~(\ref{fig:bigUniformTop}), we see the link between the
two limits $\kappa/\beta\ll1$ and $\kappa/\beta\gg1$ for
$\Delta=0$. The solid (dashed) line shows $\Delta Q_{\rm atomic}^{\rm
  JCH}$ ($\Delta Q_{\rm photonic}^{\rm JCH}$).  In these two limits,
the JCH chain mimics two Heisenberg spin chains, while in the middle
range, the full JCH dynamics are realized.  With increasing
$\kappa/\beta$, the photonic (atomic) dispersion of a pulse is
constant (constant) when $\kappa/\beta\ll 1$, increases for moderate
values of $\kappa/\beta$, then becomes constant (zero) when
$\kappa/\beta\gg 1$.  Dispersion is measured at the time $T=N/4\kappa$
for a system with $N=100$ cavities initially in the state given by
Eq.~({\ref{eq:equalsuper}) (i.e.~$|1\rangle\otimes|+,1\rangle$ with
  $\Delta=0$).  The horizontal lines drawn on the figure show the
  dispersion $\Delta Q^{\rm Heis}$ ($Q^{\rm Heis} = {\rm
    diag}(1,2,\ldots,N)$) of a Heisenberg spin chain with 100 spins
  and initial state $|1\rangle$ for $J=\kappa$ (dotted line) and
  $J=2\kappa$ (dot-dashed line) along the excitation chain.  The $J =
  \kappa$ Heisenberg dispersion line matches both the photonic and
  atomic dispersion in the limit when $\kappa/\beta\ll1$.  The $J = 2
  \kappa$ Heisenberg dispersion line matches the photonic dispersion
  in the limit when $\kappa/\beta\gg1$, while the atomic dispersion in
  this limit tends to zero (as the atomic mode no longer propagates).
  
\placeFigSeven
  
The left inset of Fig.~(\ref{fig:bigUniformTop}) and
Fig.~(\ref{fig:bigUniformBottom})(a) and (b) correspond to the example
$\kappa/\beta=10^{-3}$.  The inset shows the pulse profile at the time
$T=N/4\kappa$ [with pulse approximately a quarter along the chain,
  $\mathcal{Q}_{\wedge}=N/4$, and solid (dashed) line representing the
  photonic (atomic) profile], while part (a) and (b) shows that the
evolution of the atomic and photonic modes are identical.  The middle
inset of Fig.~(\ref{fig:bigUniformTop}) and
Fig.~(\ref{fig:bigUniformBottom})(c) and (d) correspond to the example
$\kappa/\beta=10$.  The inset shows the high dispersion of the pulse
profile at the time $T=N/4\kappa$ (the dashed line shows atomic
profile), and part (b) shows the high dispersion evolution.  The right
inset of Fig.~(\ref{fig:bigUniformTop}) and
Fig.~(\ref{fig:bigUniformBottom})(e) and (f) correspond to the example
$\kappa/\beta=10^3$.  The inset shows the pulse profile at
$T=N/4\kappa$, note that the photonic profile is approximately half
way along the chain ($\mathcal{Q}_{\wedge}=N/2$), while the dashed
line shows the atomic profile, which does not propagate.  The atomic
and photonic modes remain in superposition, despite the fact that the
atomic mode does not propagate and the photonic mode does.

Consider the profile (probability distribution at an instant in time)
of a single excitation as it travels along the chain.  In
Fig.~(\ref{fig:twoHeisenbergs})(a), we display the expectation value
of the evolution of a single up spin (initially at $|1\rangle$) for a
Heisenberg spin chain with uniform coupling $J$ between each of the
100 sites.  One sees that the pulse is initially well-formed, but
later suffers from increasing dispersion of the excitation, both when
travelling through the chain, and also when reflecting from the end of
the chain.  Regardless, the speed of the pulse is given by $J$, at
least for the first reflection, where the wave speed can be
realistically interpreted.  Fig.~(\ref{fig:Q})(a) is a different way
of showing the same evolution, by plotting the expectation value and
uncertainty of the operator $Q^{\rm Heis}={\rm diag}(1,2,\ldots,N)$
highlighting the spread of the pulse with time.  While the physics of
Heisenberg spin chains is well studied
\cite{ref:fisher1989,ref:stanley}, it is useful to contrast this
behaviour with the next regime we will examine.

\section{Parabolic coupling}
\label{sec:parabolic}

One way to control the dispersion of the pulse is to define a
non-uniform distribution of couplings between cavities.  One such
possibility is to choose a regime of parabolic couplings
\cite{ref:Albaneseetal,ref:christandletal}, where the coupling between
cavity $i$ and cavity $i+1$ is given by $\sqrt{i (N-i)}$, so that the
adjacency matrix becomes

\begin{equation}
\label{eq:adjacencyParabolic}
A_{ij} = \left\{ \begin{array}{ll}
\sqrt{j (N-j)} & i-j=1\\
\sqrt{i (N-i)} & j-i=1\\
0 & \mbox{otherwise}
\end{array}\right..
\end{equation}.

Thus the couplings will be symmetric around the central cavity (or
between the two central cavities), with strongest coupling at the
center of the chain and weakest coupling at the ends.  A spin chain of
length $N$ with this coupling can be mapped to a single spin
$s=(N-1)/2$ particle, placed in a magnetic field in the $x$
direction. This system provides the physical insight for why this
coupling is dispersion-free \cite{ref:Albaneseetal}.  The eigenvalues
of this matrix are $E_k = N - 1 - 2k$ and the eigenvectors are

\begin{eqnarray}
|k\rangle =&& \sum_{Q=1}^N\sqrt{\frac{(1-N)_k (N-1)!}{(-1)^k k!2^{N-1}(Q-1)!(N-Q)!}}\times\nonumber\\
&& K_k (Q-1,\frac{1}{2},N-1) |Q\rangle,
\end{eqnarray}
where the Pochhammer symbol $(N)_k$ is defined as

\begin{equation}
(N)_k = (N) (N+1)\ldots (N+k-1),\; k=1,2,3,\ldots
\end{equation}
and the Krawtchouk polynomial $K$ \cite{ref:atakishievetal,
  ref:orthogonalpolynoms} is related to the hypergeometric function
$F$ by

\begin{equation}
K_k(l,p,N) = \,_2F_1 \left.\left(\substack{-k,-l\\-N}\right|p^{-1}\right).
\end{equation}

Now we examine the behaviour of the JCH chain under this parabolic
coupling scheme.  We find that, again in the basis given by
$\{|k\rangle \otimes |g,1\rangle$, $|k\rangle \otimes|e,0\rangle\}$,
the Hamiltonian is a block diagonal matrix, with the $k$th block
appearing as

\begin{equation}
\mathcal{H}^{\rm JCH}_{\rm 1exc}(k) =
\left(\begin{array}{cc}
\frac{\Delta}{2} - \kappa (N-1-2k) & \beta\\
\beta & -\Delta/2
\end{array}
\right).
\end{equation}

The eigenvalues of the full Hamiltonian (\ref{eq:JCH1exc}) in the
parabolic coupling case are

\begin{eqnarray}
E^k_{\pm} &= &\frac{1}{2} \Big\{ \kappa (2k-N+1) \nonumber\\
&&\pm \sqrt{[\Delta + \kappa (2k-N+1) ]^2+4\beta^2}\Big\}
\end{eqnarray}
and the eigenvectors are

\begin{equation}
|\pm,k\rangle = \frac{(\Delta + 2 E_{\pm}^k) |g,1\rangle + 2\beta |e,0\rangle}{\sqrt{(\Delta+2 E^k_{\pm})^2 + 4\beta^2}}\otimes |k\rangle.
\end{equation}

We find the matching that occurs in limiting cases for the parabolic
JCH system, occurs in exactly the same way as the uniform coupling
case.  That is, when $\Delta = 0$ and $\kappa\ll\beta$, $J=\kappa$ for
both the atomic and photonic parts. When $\Delta = 0$ and
$\kappa\gg\beta$, $J=2\kappa$ for the photonic mode and the atomic
mode does not propagate. Figs.~(\ref{fig:bigParabolicBottom}) and
(\ref{fig:bigParabolicTop}) show very similar results to
Figs.~(\ref{fig:bigUniformBottom}) and (\ref{fig:bigUniformTop}), with
only one significant difference (apart from the dispersion-free
pulses), that the uncertainty in position for moderate values of
$\kappa/\beta$ is much larger than the uniform coupling case.

\section{Gaussian pulses}
\label{sec:gaussian}

\placeFigEight
\placeFigNine
\placeFigTen

\placeFigEleven

As an alternative to modifying the coupling profile, we can consider
uniform coupling and a Gaussian wave packet as our initial state.  In
this case, the momentum distribution of the pulse is well defined (and
narrow) which allows the excitation to travel down the chain with
minimal increase in dispersion \cite{ref:shenfanPRL, ref:shenfanOL}.
We therefore choose an appropriate initial state

\begin{equation}
|\psi^{(k,s,Q_c)}(t=0)\rangle = \mathcal{N}\sum_{Q=1}^N e^{-\frac{(Q-Q_c)^2}{2s^2}}e^{-i k Q} |Q\rangle,
\label{eq:gaussianHeis}
\end{equation}
for a Heisenberg spin chain, or

\begin{equation}
\begin{split}
&|\psi^{(k,s,Q_c)}(t=0)\rangle \\
&= \mathcal{N}\sum_{Q=1}^N e^{-\frac{(Q-Q_c)^2}{2s^2}}e^{-i k Q} |Q\rangle \otimes (|g,1\rangle + |e,0\rangle)/\sqrt{2},
\label{eq:gaussianJCH}
\end{split}
\end{equation}
for a JCH chain, where $\mathcal{N}$ is the normalization, $k$ is the
wave number, $Q_c$ is the center of the pulse, $s$ is the width of the
pulse, and $Q$ denotes cavity number, with $Q=1,\ldots,N$.  In this
paper we choose $Q_c=N/2$, $s=N/10$, such that the pulse is initiated
at the center of the chain with a width approximately 2/10 times the
length of the chain.  The value $k=\pi/2+n\pi$, where $n$ is an
integer, produces dispersion-free evolution of the pulse.

Fig.~(\ref{fig:bigGaussianBottom}) shows the evolution of a Gaussian
pulse for three different values of $\kappa/\beta$, in analogy to
Fig.~(\ref{fig:bigUniformBottom}) and (\ref{fig:bigParabolicBottom}).
The white dashed lines correspond to a triangle wave similar to
Eq.~(\ref{eq:triangle}), but with a phase shift by $\pi/2$.  The phase
shift is necessary as we have initiated the Gaussian pulse in the
center of the chain, to avoid boundary effects (contrary to the first
uniform coupling case, where the boundary effects were essential for
the motion of the pulse).  The dashed line is given by

\begin{equation}
\mathcal{Q}^{\rm Gaussian}_{\wedge} = \frac{N-1}{\pi}
\arcsin\left[\sin\left(\frac{J\pi t}{N} \right)\right] +
\frac{N+1}{2}.
\label{eq:triangleGaussian}
\end{equation}

In the limits $\kappa/\beta \ll 1$ and $\kappa/\beta \gg 1$, the
Gaussian pulse with the choice of $k=\pi/2$ evolves under simple
translation motion, interrupted by reflection off the boundaries.
This translational motion, as well as the interference pattern as it
reflects off the boundary, mimics exactly the propagation of a
Gaussian wave packet under the one-dimensional Schr\"odinger equation
\cite{ref:anatomy, ref:shenfanPRL, ref:shenfanOL}.

Fig.~(\ref{fig:bigGaussianTop}) shows the dispersion of the pulse as a
function of $\kappa/\beta$, in analogy to
Fig.~(\ref{fig:bigUniformTop}) and (\ref{fig:bigParabolicTop}).  Note
how this figure is much simpler than the earlier two, as the
dispersion is a constant throughout the evolution of the pulse,
provided the pulse is not interacting with the boundaries.  As such,
only one horizontal line to indicate dispersion of a corresponding
Heisenberg spin chain is necessary.  Also, the dispersion of the
photonic mode is equal to the dispersion of the atomic mode.


By comparing Fig.~(\ref{fig:twoHeisenbergs})(a) (uniform coupling)
with (\ref{fig:twoHeisenbergs})(b) (parabolic coupling) and
Fig.~(\ref{fig:twoHeisenbergs})(c) (uniform coupling, Gaussian pulse)
we can indeed see that the pulse in the uniform chain has more
dispersion than that of the parabolic chain and the uniform chain with
an initial Gaussian pulse, the dispersion is manifest as a number of
faint lines parallel to the main wavefront.  Further, in
Fig.~(\ref{fig:Q}) we have plotted the expectation value of the
position $\langle Q\rangle$ (solid lines) and $\Delta Q$ (dashed
lines), for both the uniform coupling (a), parabolic coupling (b) and
uniform coupling with Gaussian pulse (c) cases.  In fact, the
expectation value of the position $Q$ in the parabolic chain is of
this simple form:

\begin{equation}
\label{eq:Qev}
\langle Q(t)\rangle = \frac{1}{2} [ N + 1 - (N-1) \cos J t].
\end{equation}
Hence the period of oscillation is $2\pi/J$, which does not depend on
the number of cavities $N$.  For the coupling profile, we find that
the pulse (while more dispersed) is approximately Gaussian shaped in
space.  This can be seen by examining
Fig.~(\ref{fig:pulseProfiles})(b), which shows the pulse profile at
fixed instants in time.  In comparison,
Fig.~(\ref{fig:pulseProfiles})(a) has much higher dispersion, while
Fig.~(\ref{fig:pulseProfiles})(c) also has low dispersion, which is
largely constant and only reduces slightly as the pulse is reflected
at the boundaries of the chain.

\section{Large detuning limit}
\label{sec:largedetuning}

We now consider the case where the magnitude of the detuning is much
larger than the other energy scales of the system, that is,
$\kappa,\beta\ll|\Delta|$.  The effect of increasing detuning is to
\emph{decrease} the atom-photon coupling, thereby approximating a
Bose-Hubbard system.  We are therefore interested in any non-trivial
effects which appear in this limit.

We now shift again to an interaction picture, where states of the system $|\psi\rangle$ are transformed to 

\begin{equation}
|\xi\rangle =
e^{i (\Delta I\otimes Z/2 + \beta I\otimes X ) t}|\psi\rangle.
\end{equation}
Under this shift, the Hamiltonian becomes

\begin{equation}
\label{eq:DeltaHuge}
\begin{split}
H = -\kappa A \otimes \Big[& \frac{I+Z}{2}
  \left(1-\frac{2\beta^2}{\Delta^2+4\beta^2 }\right)\\
 + &\frac{I-Z}{2}
  \frac{2\beta^2}{\Delta^2+4\beta^2 }+ \frac{\Delta\beta }{\Delta^2+4\beta^2} X\Big].
\end{split}
\end{equation}

In this equation, the coefficient of $(I+Z)/2$ is related to the speed
of the atomic component, and the coefficient of $(I-Z)/2$ is related
to the speed of the photonic component.  We will discuss the effect of
the $X$ term shortly.

To explore the dynamics of this system, we again start with the equal
superposition initial state as given in Eq.~(\ref{eq:equalsuper}).
For example, the uniform coupling case with $\Delta=10^3\beta$,
$\kappa=\beta$ is shown in Fig.~(\ref{fig:differentSpeeds})(a).  By
again comparing to the Heisenberg spin chain Hamiltonian,
Eq.~(\ref{eq:Heis1up}), we see now that the photonic excitation
corresponds to a Heisenberg spin chain with speed $J=2\kappa
[1-2\beta^2/(4\beta^2 + \Delta^2)]$, and the atomic excitation
corresponds to a Heisenberg spin chain with speed
$J=2\kappa\beta^2/(4\beta^2 + \Delta^2)$. In this regime, the atomic
and photonic excitations travel at two completely different speeds:
the photonic mode travels much faster than the atomic mode, although
the integrity of each mode is preserved.  We also can see in
Fig.~(\ref{fig:differentSpeeds}) that this splitting of the system
into two separate modes happens for both parabolic coupling (b) and an
initial Gaussian pulse (c).  This splitting is an observation which
can be predicted directly from the $\kappa A$ independent form of the
various components of Eq.~(\ref{eq:DeltaHuge}).  Note that the final
term of Eq.~(\ref{eq:DeltaHuge}) is significantly different than both
the time scale of the photonic propagation ($\approx2\kappa$) and the
time scale of the atomic propagation
($\approx2\kappa\beta^2/\Delta^2$).  As such it does not contribute
significantly to either mode of propagation in the very large detuning
limit.

If the detuning is reduced such that this separation of time scales is
not so strong, we must ask the question what is the effect of this
final $X$ term in Eq.~(\ref{eq:DeltaHuge})?  To answer this, we write
it out in the original basis of atoms and photons on neighboring sites
$j$ and $j+1$,

\begin{equation}
\label{eq:Xterm}
X_{j,j+1} = \sigma_j^+ a_{j+1} + \sigma_j^- a_{j+1}^\dag + \sigma_{j+1}^+ a_j + \sigma_{j+1}^- a_j^\dag,
\end{equation}
in which we see that it is an effective JC type coupling between neighboring atoms and photons.  The effect of this term on the propagating atom and photon chains will be zero at first order as it couples \emph{between} chains.  The second order effect is not zero, as 

\begin{equation}
\begin{split}
\label{eq:XtermCommut}
[&X_{j,j+1}, X_{j+1,j+2}] =\\ & - (\sigma_j^+ \sigma_{j+2}^- -
\sigma_{j+2}^+ \sigma_j^- + a_j^\dag a_{j+2} - a_{j+2}^\dag a_j),
\end{split}
\end{equation}
giving an effective next-nearest-neighbor coupling (assuming the one
excitation subspace).  The strength of this correction is given by the
square of the $A\otimes X$ coefficient in Eq.~(\ref{eq:DeltaHuge}) and
therefore vanishes as the detuning is increased.  In the limit where
mixing effects are visible due to this additional term, we also
observe asymmetric behavior with respect to the sign of the detuning,
similar to that seen previously in a two cavity system
\cite{ref:irish}.  It should be noted that here, we always initialize
the system in an equal superposition of atomic and photonic states,
whereas in previous work, the detuning asymmetry is accentuated by an
initial state which is always an eigenstate of the JC Hamiltonian.

\section{Conclusions}
\label{sec:conclusions}

In conclusion, we find that when limited to the one excitation
subspace, one can analytically solve for the evolution of the
excitation pulse.  As such, we extend previous work on the topic
\cite{ref:irish, ref:ogden} from $N=2$ to the many-cavity regime.  We
observe both localized and delocalized behavior in this system, which
points to a complex interplay between atomic and photonic degrees of
freedom, even in the single excitation limit.

We consider three natural limits, and show that the behaviour of the
atomic and photonic modes of the JCH chain can be mapped to two
independent Heisenberg spin chains.  In the limit when the detuning is
zero and the intercavity coupling is much smaller than the atom-photon
coupling, we find that the system is mapped to two Heisenberg spin
chains both with $J=\kappa$.  When the detuning is zero but
atom-photon coupling is much smaller than intercavity coupling, the
photonic mode propagates with approximate speed $J=2\kappa$ and the
atomic mode does not.

In the limit when the detuning is much larger than both the
intercavity coupling and the atom-photon coupling, the system is
mapped to two Heisenberg spin chains, with $J=2\kappa
[1-2\kappa\beta^2/(\Delta^2+4\beta^2)]$ for the photonic mode, and $J
= 2\kappa\beta^2/(\Delta^2+4\beta^2)$ for the atomic mode.  We also
derive similar analytic solutions for the case of a parabolic
variation in inter-cavity coupling, resulting in a Gaussian like wave
packet propagation along the chain, as well as an initial Gaussian
pulse in a uniform coupled chain.

The JCH system provides an interesting playground for studying
many-body physics in an atom photon context.  Given that initial
experiments will be limited in both cavity and excitation number, it
is important we understand these temporal dynamics in a range of
system size regimes.

\section{Acknowledgements}

M.M.~wishes to thank Dr.~Brendon Lovett for useful discussions.  J.H.C.~wishes to acknowledge the support of the Alexander von Humboldt
foundation.  A.D.G.~and L.C.L.H.~acknowledge the Australian Research
Council for financial support Projects No. DP0880466 and No.
DP0770715, respectively .  This work was supported by the Australian
Research Council, the Australian Government, and the US National
Security Agency (NSA) and the Army Research Office (ARO) under
contract number W911NF-08-1-0527.

\bibliography{../../thesis/papers}

\begin{thebibliography}{33}
\expandafter\ifx\csname natexlab\endcsname\relax\def\natexlab#1{#1}\fi
\expandafter\ifx\csname bibnamefont\endcsname\relax
  \def\bibnamefont#1{#1}\fi
\expandafter\ifx\csname bibfnamefont\endcsname\relax
  \def\bibfnamefont#1{#1}\fi
\expandafter\ifx\csname citenamefont\endcsname\relax
  \def\citenamefont#1{#1}\fi
\expandafter\ifx\csname url\endcsname\relax
  \def\url#1{\texttt{#1}}\fi
\expandafter\ifx\csname urlprefix\endcsname\relax\def\urlprefix{URL }\fi
\providecommand{\bibinfo}[2]{#2}
\providecommand{\eprint}[2][]{\url{#2}}

\bibitem[{\citenamefont{Bloch}(2005)}]{ref:bloch}
\bibinfo{author}{\bibfnamefont{I.}~\bibnamefont{Bloch}}, \bibinfo{journal}{Nat.
  Phys.} \textbf{\bibinfo{volume}{1}}, \bibinfo{pages}{23}
  (\bibinfo{year}{2005}).

\bibitem[{\citenamefont{Balili et~al.}(2007)\citenamefont{Balili, Hartwell,
  Snoke, Pfeiffer, and West}}]{ref:balili}
\bibinfo{author}{\bibfnamefont{R.}~\bibnamefont{Balili}},
  \bibinfo{author}{\bibfnamefont{V.}~\bibnamefont{Hartwell}},
  \bibinfo{author}{\bibfnamefont{D.}~\bibnamefont{Snoke}},
  \bibinfo{author}{\bibfnamefont{L.}~\bibnamefont{Pfeiffer}}, \bibnamefont{and}
  \bibinfo{author}{\bibfnamefont{K.}~\bibnamefont{West}},
  \bibinfo{journal}{Science} \textbf{\bibinfo{volume}{316}},
  \bibinfo{pages}{1007} (\bibinfo{year}{2007}).

\bibitem[{\citenamefont{Hartmann et~al.}(2006)\citenamefont{Hartmann, {F. G. S.
  L. Brand\~ao}, and Plenio}}]{ref:hartmannNP}
\bibinfo{author}{\bibfnamefont{M.~J.} \bibnamefont{Hartmann}},
  \bibinfo{author}{\bibnamefont{{F. G. S. L. Brand\~ao}}}, \bibnamefont{and}
  \bibinfo{author}{\bibfnamefont{M.~B.} \bibnamefont{Plenio}},
  \bibinfo{journal}{Nat. Phys.} \textbf{\bibinfo{volume}{2}},
  \bibinfo{pages}{849} (\bibinfo{year}{2006}).

\bibitem[{\citenamefont{Greentree et~al.}(2006)\citenamefont{Greentree, Tahan,
  Cole, and Hollenberg}}]{ref:GTCH}
\bibinfo{author}{\bibfnamefont{A.~D.} \bibnamefont{Greentree}},
  \bibinfo{author}{\bibfnamefont{C.}~\bibnamefont{Tahan}},
  \bibinfo{author}{\bibfnamefont{J.~H.} \bibnamefont{Cole}}, \bibnamefont{and}
  \bibinfo{author}{\bibfnamefont{L.~C.~L.} \bibnamefont{Hollenberg}},
  \bibinfo{journal}{Nat. Phys.} \textbf{\bibinfo{volume}{2}},
  \bibinfo{pages}{856} (\bibinfo{year}{2006}).

\bibitem[{\citenamefont{Angelakis et~al.}(2007)\citenamefont{Angelakis, Santos,
  and Bose}}]{ref:angelakis07}
\bibinfo{author}{\bibfnamefont{D.~G.} \bibnamefont{Angelakis}},
  \bibinfo{author}{\bibfnamefont{M.~F.} \bibnamefont{Santos}},
  \bibnamefont{and} \bibinfo{author}{\bibfnamefont{S.}~\bibnamefont{Bose}},
  \bibinfo{journal}{Phys. Rev. A} \textbf{\bibinfo{volume}{76}},
  \bibinfo{pages}{031805(R)} (\bibinfo{year}{2007}).

\bibitem[{\citenamefont{Rossini and Fazio}(2007)}]{ref:pointylobes}
\bibinfo{author}{\bibfnamefont{D.}~\bibnamefont{Rossini}} \bibnamefont{and}
  \bibinfo{author}{\bibfnamefont{R.}~\bibnamefont{Fazio}},
  \bibinfo{journal}{Phys. Rev. Lett.} \textbf{\bibinfo{volume}{99}},
  \bibinfo{pages}{186401} (\bibinfo{year}{2007}).

\bibitem[{\citenamefont{Bose et~al.}(2007)\citenamefont{Bose, Angelakis, and
  Burgath}}]{ref:BoseAngelakisBurgath}
\bibinfo{author}{\bibfnamefont{S.}~\bibnamefont{Bose}},
  \bibinfo{author}{\bibfnamefont{D.~G.} \bibnamefont{Angelakis}},
  \bibnamefont{and} \bibinfo{author}{\bibfnamefont{D.}~\bibnamefont{Burgath}},
  \bibinfo{journal}{Journ. of Mod. Opt.} \textbf{\bibinfo{volume}{54}},
  \bibinfo{pages}{2307} (\bibinfo{year}{2007}).

\bibitem[{\citenamefont{Irish et~al.}(2008)\citenamefont{Irish, Ogden, and
  Kim}}]{ref:irish}
\bibinfo{author}{\bibfnamefont{E.~K.} \bibnamefont{Irish}},
  \bibinfo{author}{\bibfnamefont{C.~D.} \bibnamefont{Ogden}}, \bibnamefont{and}
  \bibinfo{author}{\bibfnamefont{M.~S.} \bibnamefont{Kim}},
  \bibinfo{journal}{Phys. Rev. A} \textbf{\bibinfo{volume}{77}},
  \bibinfo{pages}{033801} (\bibinfo{year}{2008}).

\bibitem[{\citenamefont{Ogden et~al.}(2008)\citenamefont{Ogden, Irish, and
  Kim}}]{ref:ogden}
\bibinfo{author}{\bibfnamefont{C.~D.} \bibnamefont{Ogden}},
  \bibinfo{author}{\bibfnamefont{E.~K.} \bibnamefont{Irish}}, \bibnamefont{and}
  \bibinfo{author}{\bibfnamefont{M.~S.} \bibnamefont{Kim}},
  \bibinfo{journal}{Phys. Rev. A} \textbf{\bibinfo{volume}{78}},
  \bibinfo{pages}{063805} (\bibinfo{year}{2008}).

\bibitem[{\citenamefont{Koch and Hur}(2009)}]{ref:koch}
\bibinfo{author}{\bibfnamefont{J.}~\bibnamefont{Koch}} \bibnamefont{and}
  \bibinfo{author}{\bibfnamefont{K.~L.} \bibnamefont{Hur}},
  \bibinfo{journal}{arXiv:0905.4005}  (\bibinfo{year}{2009}).

\bibitem[{\citenamefont{Na et~al.}(2008)\citenamefont{Na, Utsunomiya, Tian, and
  Yamamoto}}]{ref:na2yamamoto}
\bibinfo{author}{\bibfnamefont{N.} \bibnamefont{Na}},
  \bibinfo{author}{\bibfnamefont{S.}~\bibnamefont{Utsunomiya}},
  \bibinfo{author}{\bibfnamefont{L.}~\bibnamefont{Tian}}, \bibnamefont{and}
  \bibinfo{author}{\bibfnamefont{Y.}~\bibnamefont{Yamamoto}},
  \bibinfo{journal}{Phys. Rev. A} \textbf{\bibinfo{volume}{77}},
  \bibinfo{pages}{031803(R)} (\bibinfo{year}{2008}).

\bibitem[{\citenamefont{Grochol}(2009)}]{ref:grochol}
\bibinfo{author}{\bibfnamefont{M.}~\bibnamefont{Grochol}},
  \bibinfo{journal}{Phys. Rev. B} \textbf{\bibinfo{volume}{79}},
  \bibinfo{pages}{205306} (\bibinfo{year}{2009}).

\bibitem[{\citenamefont{Wallraff et~al.}(2004)\citenamefont{Wallraff, Schuster,
  Frunzio, Huang, Majer, Kumar, Girvin, and Schoelkopf}}]{ref:wallraff}
\bibinfo{author}{\bibfnamefont{A.}~\bibnamefont{Wallraff}},
  \bibinfo{author}{\bibfnamefont{D.~I.} \bibnamefont{Schuster}},
  \bibinfo{author}{\bibfnamefont{L.}~\bibnamefont{Frunzio}},
  \bibinfo{author}{\bibfnamefont{R.~S.} \bibnamefont{Huang}},
  \bibinfo{author}{\bibfnamefont{J.}~\bibnamefont{Majer}},
  \bibinfo{author}{\bibfnamefont{S.}~\bibnamefont{Kumar}},
  \bibinfo{author}{\bibfnamefont{S.~M.} \bibnamefont{Girvin}},
  \bibnamefont{and} \bibinfo{author}{\bibfnamefont{R.~J.}
  \bibnamefont{Schoelkopf}}, \bibinfo{journal}{Nature}
  \textbf{\bibinfo{volume}{431}}, \bibinfo{pages}{162} (\bibinfo{year}{2004}).

\bibitem[{\citenamefont{Ji et~al.}(2009)\citenamefont{Ji, Sun, Xie, and
  Liu}}]{ref:jisunxieliu}
\bibinfo{author}{\bibfnamefont{A.-C.} \bibnamefont{Ji}},
  \bibinfo{author}{\bibfnamefont{Q.}~\bibnamefont{Sun}},
  \bibinfo{author}{\bibfnamefont{X.~C.} \bibnamefont{Xie}}, \bibnamefont{and}
  \bibinfo{author}{\bibfnamefont{W.~M.} \bibnamefont{Liu}},
  \bibinfo{journal}{Phys. Rev. Lett.} \textbf{\bibinfo{volume}{102}},
  \bibinfo{pages}{023602} (\bibinfo{year}{2009}).

\bibitem[{\citenamefont{Gerace et~al.}(2009)\citenamefont{Gerace, T\"{u}reci,
  Imamoglu, Giovannetti, and Fazio}}]{ref:gerace}
\bibinfo{author}{\bibfnamefont{D.}~\bibnamefont{Gerace}},
  \bibinfo{author}{\bibfnamefont{H.}~\bibnamefont{T\"{u}reci}},
  \bibinfo{author}{\bibfnamefont{A.}~\bibnamefont{Imamoglu}},
  \bibinfo{author}{\bibfnamefont{V.}~\bibnamefont{Giovannetti}},
  \bibnamefont{and} \bibinfo{author}{\bibfnamefont{R.}~\bibnamefont{Fazio}},
  \bibinfo{journal}{Nat. Phys.} \textbf{\bibinfo{volume}{5}},
  \bibinfo{pages}{281} (\bibinfo{year}{2009}).

\bibitem[{\citenamefont{Trupke et~al.}(2005)\citenamefont{Trupke, Hinds,
  Eriksson, Curtis, Moktadir, Kukharenka, and Kraft}}]{ref:trupke}
\bibinfo{author}{\bibfnamefont{M.}~\bibnamefont{Trupke}},
  \bibinfo{author}{\bibfnamefont{E.~A.} \bibnamefont{Hinds}},
  \bibinfo{author}{\bibfnamefont{S.}~\bibnamefont{Eriksson}},
  \bibinfo{author}{\bibfnamefont{E.~A.} \bibnamefont{Curtis}},
  \bibinfo{author}{\bibfnamefont{Z.}~\bibnamefont{Moktadir}},
  \bibinfo{author}{\bibfnamefont{E.}~\bibnamefont{Kukharenka}},
  \bibnamefont{and} \bibinfo{author}{\bibfnamefont{M.}~\bibnamefont{Kraft}},
  \bibinfo{journal}{Appl. Phys. Lett.} \textbf{\bibinfo{volume}{87}},
  \bibinfo{pages}{211106} (\bibinfo{year}{2005}).

\bibitem[{\citenamefont{Shen and Fan}(2005{\natexlab{a}})}]{ref:shenfanPRL}
\bibinfo{author}{\bibfnamefont{J.-T.} \bibnamefont{Shen}} \bibnamefont{and}
  \bibinfo{author}{\bibfnamefont{S.}~\bibnamefont{Fan}},
  \bibinfo{journal}{Phys. Rev. Lett.} \textbf{\bibinfo{volume}{95}},
  \bibinfo{pages}{213001} (\bibinfo{year}{2005}{\natexlab{a}}).

\bibitem[{\citenamefont{Shen and Fan}(2005{\natexlab{b}})}]{ref:shenfanOL}
\bibinfo{author}{\bibfnamefont{J.-T.} \bibnamefont{Shen}} \bibnamefont{and}
  \bibinfo{author}{\bibfnamefont{S.}~\bibnamefont{Fan}},
  \bibinfo{journal}{Optics Letters} \textbf{\bibinfo{volume}{30}},
  \bibinfo{pages}{2001} (\bibinfo{year}{2005}{\natexlab{b}}).

\bibitem[{\citenamefont{Makin et~al.}(2008)\citenamefont{Makin, Cole, Tahan,
  Hollenberg, and Greentree}}]{ref:phasetransitions}
\bibinfo{author}{\bibfnamefont{M.~I.} \bibnamefont{Makin}},
  \bibinfo{author}{\bibfnamefont{J.~H.} \bibnamefont{Cole}},
  \bibinfo{author}{\bibfnamefont{C.}~\bibnamefont{Tahan}},
  \bibinfo{author}{\bibfnamefont{L.~C.~L.} \bibnamefont{Hollenberg}},
  \bibnamefont{and} \bibinfo{author}{\bibfnamefont{A.~D.}
  \bibnamefont{Greentree}}, \bibinfo{journal}{Phys. Rev. A}
  \textbf{\bibinfo{volume}{77}}, \bibinfo{pages}{053819}
  (\bibinfo{year}{2008}).

\bibitem[{\citenamefont{Paternostro et~al.}(2009)\citenamefont{Paternostro,
  Agarwal, and Kim}}]{ref:kimsoliton}
\bibinfo{author}{\bibfnamefont{M.}~\bibnamefont{Paternostro}},
  \bibinfo{author}{\bibfnamefont{G.~S.} \bibnamefont{Agarwal}},
  \bibnamefont{and} \bibinfo{author}{\bibfnamefont{M.~S.} \bibnamefont{Kim}},
  \bibinfo{journal}{New J. Phys.} \textbf{\bibinfo{volume}{11}},
  \bibinfo{pages}{013059} (\bibinfo{year}{2009}).

\bibitem[{\citenamefont{Lu et~al.}(2009)\citenamefont{Lu, Zhou, Kuang, and
  Sun}}]{ref:sunsoliton}
\bibinfo{author}{\bibfnamefont{J.}~\bibnamefont{Lu}},
  \bibinfo{author}{\bibfnamefont{L.}~\bibnamefont{Zhou}},
  \bibinfo{author}{\bibfnamefont{L.-M.} \bibnamefont{Kuang}}, \bibnamefont{and}
  \bibinfo{author}{\bibfnamefont{C.P}~\bibnamefont{Sun}},
  \bibinfo{journal}{Phys. Rev. E} \textbf{\bibinfo{volume}{79}},
  \bibinfo{pages}{016606} (\bibinfo{year}{2009}).

\bibitem[{\citenamefont{Altug and Vu\u{c}kovi\'{c}}(2005)}]{ref:altug}
\bibinfo{author}{\bibfnamefont{H.}~\bibnamefont{Altug}} \bibnamefont{and}
  \bibinfo{author}{\bibfnamefont{J.}~\bibnamefont{Vu\u{c}kovi\'{c}}},
  \bibinfo{journal}{Appl. Phys. Lett.} \textbf{\bibinfo{volume}{86}},
  \bibinfo{pages}{111102} (\bibinfo{year}{2005}).

\bibitem[{\citenamefont{Mattis}(1993)}]{ref:mattis}
\bibinfo{editor}{\bibfnamefont{D.~C.} \bibnamefont{Mattis}}, ed.,
  \emph{\bibinfo{title}{The Many-Body Problem}} (\bibinfo{publisher}{World
  Scientific}, \bibinfo{address}{Singapore}, \bibinfo{year}{1993}).

\bibitem[{\citenamefont{Sachdev}(1999)}]{ref:sachdev}
\bibinfo{author}{\bibfnamefont{S.}~\bibnamefont{Sachdev}},
  \emph{\bibinfo{title}{Quantum phase transitions}}
  (\bibinfo{publisher}{Cambridge University Press},
  \bibinfo{address}{Cambridge}, \bibinfo{year}{1999}).

\bibitem[{\citenamefont{Jaynes and Cummings}(1963)}]{ref:jaynescummings}
\bibinfo{author}{\bibfnamefont{E.~T.} \bibnamefont{Jaynes}} \bibnamefont{and}
  \bibinfo{author}{\bibfnamefont{F.~W.} \bibnamefont{Cummings}}, in
  \emph{\bibinfo{booktitle}{Proc. of the IEEE}} (\bibinfo{year}{1963}),
  vol.~\bibinfo{volume}{51}, p.~\bibinfo{pages}{89}.

\bibitem[{\citenamefont{Christandl et~al.}(2004)\citenamefont{Christandl,
  Datta, Ekert, and Landahl}}]{ref:christandletal}
\bibinfo{author}{\bibfnamefont{M.}~\bibnamefont{Christandl}},
  \bibinfo{author}{\bibfnamefont{N.}~\bibnamefont{Datta}},
  \bibinfo{author}{\bibfnamefont{A.}~\bibnamefont{Ekert}}, \bibnamefont{and}
  \bibinfo{author}{\bibfnamefont{A.~J.} \bibnamefont{Landahl}},
  \bibinfo{journal}{Phys. Rev. Lett.} \textbf{\bibinfo{volume}{92}},
  \bibinfo{pages}{187902} (\bibinfo{year}{2004}).

\bibitem[{\citenamefont{Nielsen and Chuang}(2000)}]{ref:mikeandike}
\bibinfo{author}{\bibfnamefont{M.}~\bibnamefont{Nielsen}} \bibnamefont{and}
  \bibinfo{author}{\bibfnamefont{I.}~\bibnamefont{Chuang}},
  \emph{\bibinfo{title}{Quantum Computation and Information}}
  (\bibinfo{publisher}{Cambridge University Press}, \bibinfo{year}{2000}),
  \bibinfo{edition}{2nd} ed.

\bibitem[{\citenamefont{Fisher et~al.}(1989)\citenamefont{Fisher, Weichman,
  Grinstein, and Fisher}}]{ref:fisher1989}
\bibinfo{author}{\bibfnamefont{M.~P.~A.} \bibnamefont{Fisher}},
  \bibinfo{author}{\bibfnamefont{P.~B.} \bibnamefont{Weichman}},
  \bibinfo{author}{\bibfnamefont{G.}~\bibnamefont{Grinstein}},
  \bibnamefont{and} \bibinfo{author}{\bibfnamefont{D.~S.}
  \bibnamefont{Fisher}}, \bibinfo{journal}{Phys. Rev. B}
  \textbf{\bibinfo{volume}{40}}, \bibinfo{pages}{546} (\bibinfo{year}{1989}).

\bibitem[{\citenamefont{Stanley}(1969)}]{ref:stanley}
\bibinfo{author}{\bibfnamefont{H.~E.} \bibnamefont{Stanley}},
  \bibinfo{journal}{Phys. Rev.} \textbf{\bibinfo{volume}{179}},
  \bibinfo{pages}{570} (\bibinfo{year}{1969}).

\bibitem[{\citenamefont{Albanese et~al.}(2004)\citenamefont{Albanese,
  Christandl, Datta, and Ekert}}]{ref:Albaneseetal}
\bibinfo{author}{\bibfnamefont{C.}~\bibnamefont{Albanese}},
  \bibinfo{author}{\bibfnamefont{M.}~\bibnamefont{Christandl}},
  \bibinfo{author}{\bibfnamefont{N.}~\bibnamefont{Datta}}, \bibnamefont{and}
  \bibinfo{author}{\bibfnamefont{A.}~\bibnamefont{Ekert}},
  \bibinfo{journal}{Phys. Rev. Lett.} \textbf{\bibinfo{volume}{93}},
  \bibinfo{pages}{230502} (\bibinfo{year}{2004}).

\bibitem[{\citenamefont{Atakishiev et~al.}(1998)\citenamefont{Atakishiev,
  Jafarov, Nagiyev, and Wolf}}]{ref:atakishievetal}
\bibinfo{author}{\bibfnamefont{A.~N.} \bibnamefont{Atakishiev}},
  \bibinfo{author}{\bibfnamefont{E.}~\bibnamefont{Jafarov}},
  \bibinfo{author}{\bibfnamefont{S.}~\bibnamefont{Nagiyev}}, \bibnamefont{and}
  \bibinfo{author}{\bibfnamefont{K.}~\bibnamefont{Wolf}},
  \bibinfo{journal}{Revista Mexicana de Fisica} \textbf{\bibinfo{volume}{44}},
  \bibinfo{pages}{235} (\bibinfo{year}{1998}).

\bibitem[{\citenamefont{Al-Salam}(1990)}]{ref:orthogonalpolynoms}
\bibinfo{author}{\bibfnamefont{W.}~\bibnamefont{Al-Salam}},
  \emph{\bibinfo{title}{Orthogonal Polynomials: Theory and Practice}}
  (\bibinfo{publisher}{Kluwer Academic}, \bibinfo{address}{Dordrecht},
  \bibinfo{year}{1990}).

\bibitem[{\citenamefont{Doncheski and Robinett}(1999)}]{ref:anatomy}
\bibinfo{author}{\bibfnamefont{M.~A.} \bibnamefont{Doncheski}}
  \bibnamefont{and} \bibinfo{author}{\bibfnamefont{R.~W.}
  \bibnamefont{Robinett}}, \bibinfo{journal}{Eur. J. Phys.}
  \textbf{\bibinfo{volume}{20}}, \bibinfo{pages}{29} (\bibinfo{year}{1999}).

\end{thebibliography}

\end{document}